\documentclass[aip,amsmath,amssymb,reprint]{revtex4-1}
\usepackage{stix}
\usepackage[margin=1in]{geometry}
\usepackage[utf8]{inputenc}
\usepackage{tikz}
\usetikzlibrary{shapes.geometric}
\usepackage{amsmath}
\usepackage{amsfonts}
\usepackage{mathtools}
\usepackage{subfiles}
\usepackage[export]{adjustbox}
\usepackage{ragged2e}
\usepackage{bm} 
\usepackage[caption=false]{subfig}
\usepackage[normalem]{ulem}

\newcommand{\tr}{\operatorname{Tr}}
\def\avg#1{\mathinner{\langle{#1}\rangle}}
\def\bra#1{\mathinner{\langle{#1}|}}
\def\ket#1{\mathinner{|{#1}\rangle}}
\def\braa#1{\mathinner{\langle\langle{#1}|}}
\def\kett#1{\mathinner{|{#1}\rangle\rangle}}

\newcommand{\braakett}[2]{\langle\langle #1|#2\rangle\rangle}

\begin{document}

\title{Constructing Tensor Network Influence Functionals for General Quantum Dynamics}

\author{Erika Ye}
\email{eyerik.a@gmail.com}
\affiliation{Division of Engineering and Applied Sciences, California Institute of Technology, Pasadena, USA 91125}

\author{Garnet Kin-Lic Chan}
\email{gkc1000@gmail.com}
\affiliation{Division of Chemistry and Chemical Engineering, California Institute of Technology, Pasadena, USA 91125}

\date{\today}

\begin{abstract}
    We describe an iterative formalism to compute influence functionals that describe the general quantum dynamics of a subsystem beyond the assumption of linear coupling to a quadratic bath. We use a space-time tensor network representation of the influence functional and investigate its approximability in terms of its bond dimension and time-like entanglement in the tensor network description. We study two numerical models, the spin-boson model and a model of interacting hard-core bosons in a 1D harmonic trap.
    We find that the influence functional and the intermediates involved in its construction can be efficiently approximated by low bond dimension tensor networks in certain dynamical regimes, which allows the quantum dynamics to be accurately
    computed for longer times than with direct time evolution methods. However, as one iteratively integrates out the bath, the correlations in the influence functional can first increase before decreasing, indicating that the final compressibility of the influence functional is achieved via non-trivial cancellation.
\end{abstract}

\maketitle

\section{Introduction}
Obtaining the long-time dynamics of a large quantum system is in general intractable due to the exponential scaling of Hilbert space with respect to system size and the associated exponential growth of spatial entanglement with time.
Fortunately, in many cases, one is most interested in the dynamics of observables defined on a small subset of the full system, and can thus reframe the dynamics of the observable from the viewpoint of a subsystem coupled to a bath \cite{Breuer_book}. 
If one is able to determine the dynamics of the bath and its influence on the subsystem, then the original dynamics problem is reduced to obtaining the dynamics of the subsystem \cite{Feynman_IF, Nakajima_1958, Zwanzig_1960, TCL_SB}.

The influence functional (IF) \cite{Feynman_IF} method provides an exact framework for computing the dynamics of an arbitrary bath and its interactions with the subsystem. However, the cost of computing the IF without approximation is comparable to determining the dynamics of the full system and the size of the IF scales exponentially with the number of time steps. Thus it is not usually possible to use the IF method without additional approximations.

The IF can be viewed as reweighting the path integral of the subsystem. In the case of linear coupling to a harmonic bath, Feynman and Vernon derived an analytical form of the IF % weight
\cite{Vernon_Feynman, Vernon_IF}, which takes the form of the Boltzmann weight of a complex valued Hamiltonian defined in the time direction with pairwise interactions between time points.

For many physical bath spectral densities, it is natural to assume that the pairwise time interaction is short-ranged in time, corresponding to a finite "memory" in the influence of the bath, and many numerical approximations have successfully taken advantage of this short-range temporal nature \cite{Makri_1995, Makri_anharmonic, Strathearn_QUAPI, Weiss_iterative_PI, Segal_iterative_PI, Cohen_inchworm_0, Chen_inchworm_1, NZM_TTM, NZM_TTM_SB, GQME_2003, GQME_inouteq, CohenRabani_2011, Rabani_ReducedProp, Chatterjee_overview}. For IF methods, the assumption of limited memory allows one to remove the exponential growth of cost of the quantum dynamics with simulation time, thus making long time-scale quantum dynamical simulations possible. For example, in QuAPI \cite{Makri_1995}, %, makri_2014_blip, makri_2020_smatpi},
one approximates the analytical harmonic bath IF by only including terms acting on time steps within a finite time window. 
Alternatively one can construct an ansatz for the IF; a natural choice is a matrix product state (MPS) in the time direction, which compactly encodes short-range time correlations. This approach has been used in a number of recent works, which, although they do not necessarily use the language of IFs, all proceed by constructing a compressed version of the IF \cite{Lerose_IFmatrix} or a closely related object such as the auxiliary density operator as defined in QuAPI \cite{Strathearn_TEMPO,Jorgensen_TEMPO}, the process tensor \cite{Pollock_2018,Cygorek_IFmpo}, or other variants \cite{Banuls_trTE_0, Banuls_trTE_2, Luchnikov_2019, Luchnikov_2019_ML}. 

In this work, we are interested in using the IF method for computing the dynamics of a subsystem within a general quantum system. Such a subsystem may arise as part of a larger interacting problem (in which case, the subsystem might %may
not be different from other parts of the system) or it may arise from a system-bath model. In either case, the couplings and bath cannot be assumed to be linear and quadratic respectively and thus the analytical form of the IF is not known. Instead, the IF is simply a particular integral of the space-time dynamics that must be obtained numerically. 
% 
% To do this concretely, we can use a tensor network description of the space-time dynamics. For a 1-dimensional representation of the system and bath, the tensor network is thus defined in 1+1 dimensions, and the IF corresponds to a contraction of the network to yield a final matrix product state IF defined along the time direction. 
%
To do this concretely, we can use a tensor network description of the space-time dynamics. For a 1-dimensional representation of the system and bath, the tensor network is thus defined in 1+1 dimensions.
\textcolor{black}{Traditional time evolution corresponds to contracting this network first along the time axis, and if done exactly leads to exponential cost with time.  Computing    the IF corresponds to the contraction of the network in the spatial direction to yield a final matrix product state IF defined along the time direction, and if done exactly has an exponential cost with the spatial length. A similar framework appears in the recently proposed modular path integral (MPI) procedure \cite{makri_mpi,makri_mpi_smatpi, kundu_mpi_nondiag, kundu_mpi_smatpi, makri_mpicomm, kundu_mpi_efficient}.}
% Unlike the matrices appearing in a matrix decomposition of the path integral only along the spatial direction (for example, as occurring in the modular path integral procedure \cite{makri_mpi, kundu_mpi_nondiag, kundu_mpi_smatpi, makri_mpicomm, kundu_mpi_efficient}) or along the time direction, the tensors in the space-time tensor network do not have an exponential dependence on system size or time.
However, to avoid exponential computational cost, the contractions must be performed approximately. 
\textcolor{black}{There has been some discussion regarding using filtering methods to reduce the cost of the MPI algorithm \cite{makri_mpi}. In contrast, we utilize tensor network compression and propose a tensor network contraction procedure similar to that presented by}
% The proposed approximate tensor network contraction procedure is similar to that presented by 
Banuls et al.~\cite{Banuls_trTE_0, Banuls_trTE_2} and Lerose et al.~\cite{Lerose_IFmatrix}, assuming a translationally invariant infinite problem. However, we will describe a general \textcolor{black}{tensor network} procedure to construct the IF without such assumptions and \textcolor{black}{explore the numerical feasibility of doing so to compute quantum dynamics in different regimes beyond the commonly considered linear coupling to quadratic baths.}

The paper is organized as follows. We first translate the IF into a space-time tensor network language and describe an iterative algorithm to compute it. We then investigate the compressibility of the IF and its ability to produce long-time dynamics, first for the canonical spin-boson model where the analytical IF is known, and then an interacting hard-core boson model where there is no analytical expression, which corresponds to the case of general quantum dynamics. We analyze the time-like entanglement both in the IF itself as well as the intermediates that arise as the bath is numerically integrated out. We end with a brief discussion of the implications of this work for future studies. 

\section{Theory}
\subsection{Definition of the Influence Functional}
To introduce notation, we first recall the definition of the influence functional (IF) \cite{Feynman_IF}. The influence functional describes how the path integral of a subsystem is reweighted, under the influence of dynamical coupling to a bath. To obtain an explicit form, we define a full system composed of the subsystem of interest and the coupled bath. At time $T$, we denote the subsystem density matrix by $\rho_s(s_{T})$, where $s_{T}$ is a basis for the density matrix, and the bath density matrix is analogously written as $\rho_b(b_{T})$. The basis of the full system is spanned by the product space $\{s\} \otimes \{b\}$.
The evolution of the density matrix is given by a linear operator, the Liouville operator $L$, which we partition as
$L = L_s + L_{bs}$ where $L_s$ contains the component operating only on the subsystem and $L_{bs}$ contains the component on the bath and interactions between the subsystem and bath. If we further assume the system dynamics obeys Hamiltonian evolution, then the Liouville action can be written as $L \, \cdot = [H, \cdot]$. 

Formally, $\rho_s(s_T)$ is obtained by time evolving the entire system and tracing out the bath degrees of freedom. The path integral expression, assuming a second-order Trotter decomposition of the time evolution operator into $N$ timesteps of length $\epsilon$, is
\begin{align}
     \rho_s(s_T)  &= \nonumber\\%[0.3cm]
     %\!\!\!\! 
     \tr_{b_T} \biggl[ & \sum_{s_{t_{N-1}}} \sum_{b_{t_{N-1}}}  \dots 
     \sum_{s_{t_0}} \sum_{b_{t_0}}  \nonumber \\%[0.3cm] 
    % & \!\!\!\! 
    & \braa{s_T} e^{-\frac{i}{2}L_s \epsilon} e^{-iL_{bs}\epsilon} e^{-iL_s\epsilon}  \kett{s_{t_{N-1}},b_{t_{N-1}}} \times \nonumber \\%[0.3cm]
    % & \!\!\!\!
    & \braa{s_{t_{N-1}},b_{t_{N-1}}} e^{-iL_{bs}\epsilon} e^{-iL_s\epsilon} \kett{s_{t_{N-2}},b_{t_{N-2}}} \times \dots \nonumber \\ 
    %\hspace{1.5cm}
    % & \!\!\!\! 
    \times & \braa{s_{t_1},b_{t_1}} e^{-iL_{bs}\epsilon} e^{-\frac{i}{2}L_s\epsilon}
    \kett{s_{t_0},b_{t_0}} \times \nonumber \\%[0.3cm]
      %\hspace{10cm}
    % & \!\!\!\! 
    & \braakett{s_{t_0},b_{t_0}}{\rho(s_{t_0},b_{t_0})} \, \biggr]
    \label{eq:IF_orig}
\end{align}
where $\rho(s_{t_0},b_{t_0})$ is the initial state of the system, and the double bra/ket notation indicates we are working \textcolor{black}{in Liouville space, with the density matrix being a vector in this space}. % with Liouville space vectors. 
%We can bring the trace operation inside the integral. 
For simplicity, we assume there are no correlations between the subsystem and bath initially such that $\kett{\rho(s_{t_0}, b_{t_0})} =  \kett{\rho_s(s_{t_0})}\kett{\rho_b(b_{t_0})}$. Furthermore, $L_{bs}$ is typically assumed to be diagonal in the basis $\{ s\}$ (we lift both these restrictions below). Then Eq.~\eqref{eq:IF_orig} becomes
\begin{align}
    \rho_s(s_T) &= \sum_{s_{t_{N-1}}} \, \dots \sum_{s_{t_0}} \,\, \braa{s_T} e^{-\frac{i}{2}L_s\epsilon} \kett{s_{t_{N}}} \times \nonumber \\
    & 
    \braa{s_{t_{N}}} e^{-iL_s\epsilon}\kett{s_{t_{N-1}}} \times \dots \nonumber \\
    &  \times \braa{s_{t_1}} e^{-\frac{i}{2}L_s\epsilon} \kett{s_{t_0}} \braakett{s_{t_0}}{\rho_s(s_{t_0})} \nonumber \\ & \times I(s_{t_1},s_{t_2},...,s_{t_{N}})
    \label{eq:IF_diag}
\end{align}
where $I(s_{t_1},s_{t_2},...,s_{t_{N}})$ is the influence functional
\begin{equation}
    I(\ldots)
    %s_{t_0},s_{t_1},...,s_{t_{N-1}}) 
    = \tr_{b_T}  \left[ e^{-iL_{bs}(s_{t_{N}})\epsilon}  \dots e^{-iL_{bs}(s_{t_1})\epsilon} \kett{\rho_b(b_{t_0})} \right] 
    %%%% note: correction: t_i -> t_i+1 %%%%
\end{equation}
with $L_{bs}(s) = \langle\langle s|L_{bs}|s\rangle\rangle$.
The IF assigns a complex weight to each configuration of the system path integral. Consequently, the storage of the IF grows exponentially with number of time steps $N$.  

\subsection{Generalized Influence Functional in Tensor Network Language}

\subsubsection{Influence Functional Structure}
\begin{figure}[t!]
    \includegraphics[width=\linewidth,trim={0cm 0cm 0cm 0cm}]{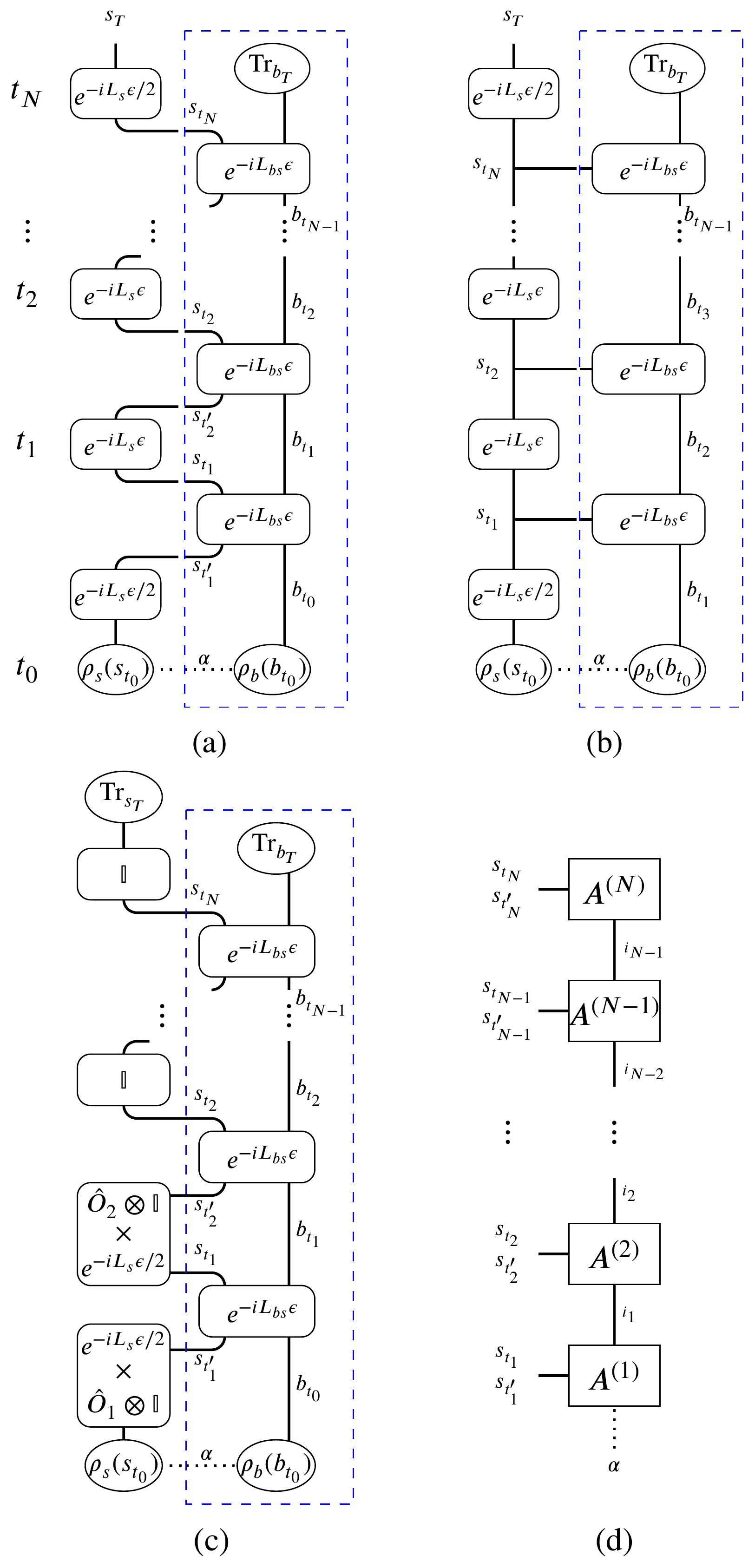}
    \caption{(a) Time evolution of $\rho(s_{t_0}) = \sum_\alpha  \rho_{s,\alpha}(s_{t_0})\otimes\rho_{b,\alpha}(s_{t_0})$ in Liouville space with second order Trotter decomposition between system and interaction dynamics. (b) The same as (a) % EY changed notation here
    but assuming $L_{bs}$ is diagonal with respect to subsystem basis. The boxed regions are the generalized and traditional definitions of the influence functional, respectively. (c) Measurement of the time-correlation $\langle \hat{O}_2(t_1)\, \hat{O}_1(t_0) \rangle_\rho $. (d) Matrix product state representation of influence functional. The labels $\{s_{t_m}\}$ and $\{b_{t_m}\}$ index the system and bath states at time step $m$, respectively. The labels $\{i_m\}$ index the virtual bonds. Lines that connect two tensors (blocks) represent tensor contraction over the labeled indices. } 
    \label{diag:IF_TN} 
\end{figure}

\textcolor{black}{Translating Eq.~\eqref{eq:IF_orig} and Eq.~\eqref{eq:IF_diag} into the tensor network language is straightforward, and is shown in diagrams Fig.~\ref{diag:IF_TN}(a) and (b), respectively. % Similar representations have been discussed in Refs.~\onlinecite{Pollock_2018, Luchnikov_2019}. 
The time evolution operators of the system and system-bath dynamics respectively appear as boxes with two and four legs in Fig.~\ref{diag:IF_TN}(a), with the legs labelled by the bra and ket basis states. The elements in the tensors are $\langle\langle{s_t} |e^{-iL_s \epsilon} | s_{t'}\rangle\rangle$ and $\langle\langle{s_t, b_t} |e^{-iL_{bs} \epsilon}| s_{t'},b_{t'}\rangle\rangle$, respectively. In the case that $L_{bs}$ is diagonal with respect to the subsystem basis states $\{ s \}$, then the elements of the system-bath time evolution operator are $\langle\langle{s_t, b_t} |e^{-iL_{bs} \epsilon}| s_{t},b_{t'}\rangle\rangle$ and can be depicted as a box with three legs labelled by $s_{t},b_{t}$, and $b_{t'}$, as in Fig.~\ref{diag:IF_TN}(b). The final trace operation over the bath degrees of freedom at the last time step can be written as a vector that is contracted with the corresponding leg of the time evolution operator tensor. 
}

\textcolor{black}{
The influence functional element $I(s_{t_1}, s_{t_2}, \ldots, s_{t_{N}})$ is the object within the blue rectangle of Fig.~\ref{diag:IF_TN}(b), obtained after performing tensor contractions over all bath tensors as denoted by the connected lines. 
Within this diagrammatic picture it is easy to depict the generalization of the influence functional to a correlated initial state. In this case, the dotted line indicates a correlated initial state with entanglement between the subsystem and bath, 
} 
\begin{align}
    \rho(s_{t_0}, b_{t_0}) = \sum_\alpha \rho_{s\alpha}(s_{t_0}) \rho_{b\alpha}(b_{t_0})
\end{align}
and the influence functional is defined with an additional index, $I(s_{t_1}, s_{t_2}, \ldots, s_{t_{N}}; \alpha)$. \textcolor{black}{Similarly, if $L_{bs}$ cannot be diagonalized in the subsystem basis $\{ s \}$ then we can generalize the influence functional to contain two subsystem indices at each intermediate time, $I(s_{t_1} s_{t'_1}, s_{t_2} s_{t'_2}, \ldots, s_{t_{N}} s_{t'_{N}})$, and it is the object within the blue rectangle of Fig.~\ref{diag:IF_TN}(a). Given the influence functional, arbitrary time-correlation functions can be computed as shown in Fig.~\ref{diag:IF_TN}(c).}

Because the influence functional has a one-dimensional structure along the time axis, it is natural to rewrite it as a matrix product state of $N$ tensors (see Fig.~\ref{diag:IF_TN}(\textcolor{black}{d})), i.e.
\begin{align}
I(s_{t_1}, s_{t_2}, \ldots, s_{t_{N}}) =& \label{eq:IF_MPS} \\%[0.1cm]
\sum_{\{ i \} } A^{(1)}_{i_1}(s_{t_1}) & A^{(2)}_{i_1,i_2}(s_{t_2}) \ldots A^{(N)}_{i_{N-1}}(s_{t_{N}}) \nonumber 
\end{align}
where $A(s_t)$ denotes a matrix of dimension $D\times D$ \textcolor{black}{for each element of the basis $s_t$}, except for $A^{(1)}(s_{t_1})$ and $A^{(N)}(s_{t_{N}})$ which are $D$ dimensional row and column vectors respectively. In MPS language, $D$ is referred to as the virtual bond dimension. Although any IF can be represented as a MPS for sufficiently large $D$, the MPS of small bond dimension naturally capture sums of exponentially decaying time-correlations along the time axis. The key system-specific questions to understand are thus (i) is the IF itself representable by an MPS of low bond dimension, in physically relevant dynamical and interaction regimes, and (ii) can the IF be constructed with manageable cost in those regimes. It is important to note that an affirmative answer to (i) does not imply an affirmative answer to (ii). 

\subsubsection{Space-Time Tensor Network Representation}

To define an approximate procedure to construct the influence functional for complex bath dynamics, we 
first write down a  space-time representation of the full system dynamics. We first assume that the bath Hilbert space is a product space over $K$ modes,
\begin{align}
\{ b \} = \{ b^1 \} \otimes \{ b^2 \} \otimes \hdots \otimes \{ b^K \}
\end{align}
We can then formally express the system density matrix at any time as a matrix product state (MPS)
\begin{align}
    \rho(s_t, b_t) = \sum_{\{ i\} } C^{(0)}_{i_0}({s_t}) \, C^{(1)}_{i_0,i_1}(b^1) \ldots C^{(K)}_{i_{K-1}}(b^K)
    \label{eq:rho_mps}
\end{align}
with a bond dimension denoted $D_\rho$.

\begin{figure}
    \includegraphics[width=\linewidth,trim={0cm 0cm 0cm 0cm}]{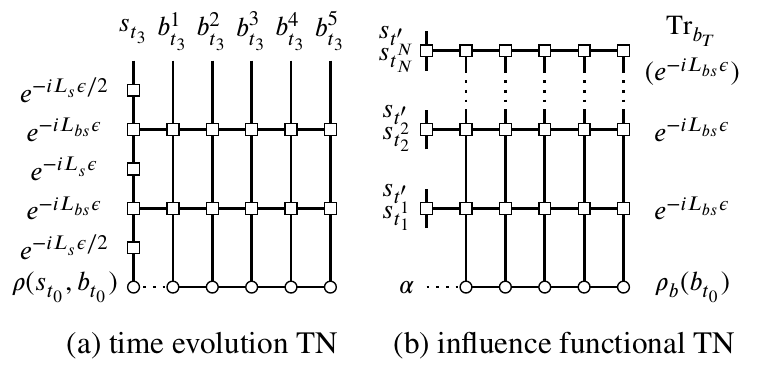}
    \caption{(a) Space-time tensor network representing time evolution of a system represented as a 1D MPS (row of circles) with second order Trotter decomposition between system (single square on left most column) and interaction dynamics (row of squares). Pictured are two time steps applied to a  system coupled to five bath sites. (b) Tensor network representation of influence functional containing $N$ time steps. The top-most row represents the time evolution operator with the trace operation applied.}
    \label{fig:spacetime}
\end{figure}

Similarly, the Liouville evolution operator can be written as a matrix product operator (MPO) \textcolor{black}{with the elements} 
\begin{align}
    \langle \langle & s_t, b_t | e^{-i L_{bs} \epsilon} | s_{t'}, b_{t'}\rangle \rangle  = 
    \label{eq:expLT_mpo}
    \\
    &\sum_{\{ i\} } M^{(0)}_{i_0}({s_t,s_{t'}}) \, M^{(1)}_{i_0,i_1}(b^1_t,b^1_{t'}) \ldots M^{(K)}_{i_{K-1}} (b^K_t, b^K_{t'}) \nonumber
\end{align}
and has a bond dimension $D_L$. Note that since the Liouville operator is assumed time-independent, $D_L$ is fixed.
\textcolor{black}{
At last the time step, the bath degrees of freedom are traced out so the bath sites are now in MPS form,  
\begin{align}
    \langle \langle s_{t_N} & |  \, \text{Tr}_{b_T} \left[ e^{-iL_{bs} \epsilon} \right] \, | \,  s_{t_N'}, b_{t_{N-1}} \rangle \rangle =
    \label{eq:expLT_mpo_trace} \\
    % &  \sum_{b_T} \langle \langle s_{t_N}, b_T | e^{-i\textcolor{black}{L_{bs}} \epsilon} | s_{t_N'}, b_{t_{N-1}} \rangle \rangle  = \nonumber \\
    &\sum_{\{ i\} } M^{(0)}_{i_0}(s_{t_N},s'_{t_N}) \tilde{M}^{(1)}_{i_0,i_1}(b^1_{t_{N-1}}) \ldots \tilde{M}^{(K)}_{i_{K-1}} (b^K_{t_{N-1}}) \nonumber
\end{align}
where $\tilde{M} = \text{Tr}_{b_T} (M)$ with $M$ as defined in Eq.~\eqref{eq:expLT_mpo}.
}

If the Hamiltonian consists of only nearest neighbor interactions, one can obtain the operator using a Trotter-Suzuki decomposition of nearest neighbour gates and then directly map them onto a matrix product operator. Otherwise, for more general interactions one can use a 4th-order Runge-Kutta expansion \cite{Ripoll_TE}. In this case, the matrix product operator can have large bond dimension, but can be compressed by allowing for truncation errors of $~\mathcal{O}(\epsilon^5)$.
In the cases studied here, the subsystem is small enough such that $e^{-i L_s \epsilon}$ can be obtained exactly.

The full time evolution of the system with $K$ bath modes and $N$ time steps thus corresponds to the two-dimensional tensor network diagram shown in Fig.~\ref{fig:spacetime}(a).
Correspondingly, the space-time representation of the influence functional is shown in Fig.~\ref{fig:spacetime}(b).

\subsubsection{Transverse Contraction Scheme}
The most common way to contract a 2D space-time tensor network is from bottom to top, i.e. in the direction of increasing time \cite{Vidal_TE,Daley_TEBD,Schollwock_MPS,White_TE,White_modifiedLTE, Schollwock_TDDMRG, Ronca_TDDMRG, Haegeman_TDVP}.
We refer to this as direct time evolution. For example, contracting the network in Fig.~\ref{fig:spacetime}(a) row by row yields the system density matrix at each time step as an MPS. 
\textcolor{black}{The cost of contracting two rows is $\mathcal{O}(K D_L^2 D_\rho^2 d_\rho^2 )$ where $d_\rho$ is the dimension of the density matrices represented at each MPS site, and $D_\rho$ is the bond dimension of the density matrix MPS. For \textit{exact} time evolution $D_\rho$ grows by a factor of $D_L$ at each time step. This means that in the worst case, $D_\rho$ grows exponentially with time.} 

However, if the time-correlations in the influence functional decay with long time, then this implies that  the influence functional ultimately can be represented by a matrix product with low bond dimension along the time axis. This suggests that a more efficient contraction strategy is to contract column wise (in the transverse direction to time).
\textcolor{black}{The cost of contracting columns together in the process of construct the final influence functional is $\mathcal{O}(N D_L^2 D_I^2 d_\rho^2 )$ where $D_I$ is the bond dimension of the column bath MPS (defined along the time direction). For \textit{exact} contraction,} $D_I$ will grow by $d_\rho$ \textcolor{black}{at each contraction step and thus in the worst case scales exponentially with $K$}. But, it need not have a dependence on the total simulation time.

In practice, \textcolor{black}{due to the exponential growth of the bond dimension,} exact contraction of the 2D tensor network (in either direction) is often too expensive. In Banuls et al~\cite{Banuls_trTE_0,Banuls_trTE_2}, the explicit transverse contraction of the 2D tensor network was avoided by assuming that the system is infinite and translationally invariant, in which case the result of the infinite contraction of columns is proportional to the maximal eigenvector of the column transfer operator.

\begin{figure*}[t]
    \centering
        \includegraphics[width=0.95\linewidth,trim={0cm 0cm 0cm 0cm}]{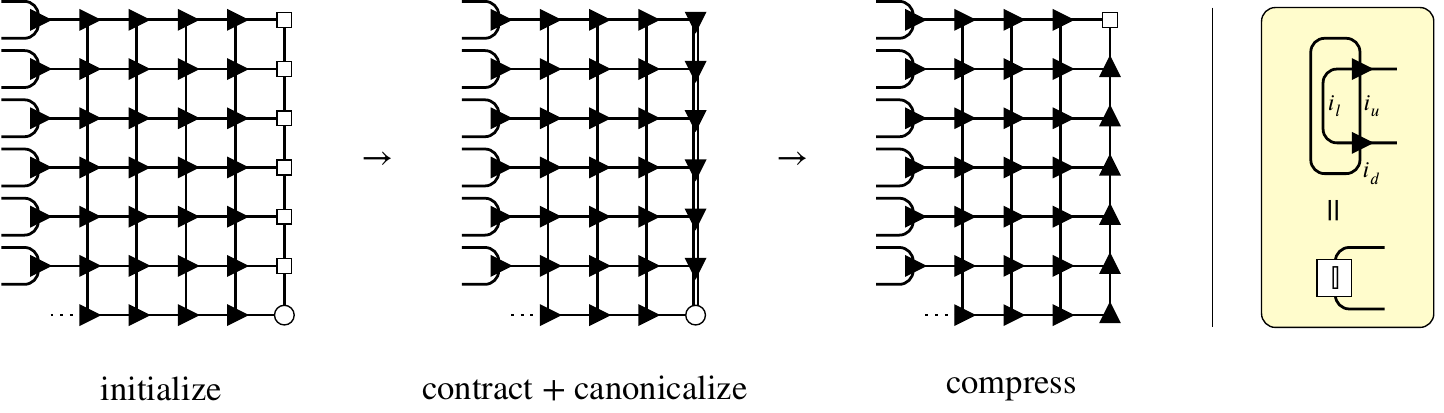}
    \caption{Iteration of transverse contraction scheme to compute IF tensor network. The time evolution operators $e^{-iL_{bs}\epsilon}$ at each time step are the rows of the grid and are each represented as an MPO, and the subsystem of interest is at the left-most site. Before contraction, we first canonicalize each row into left canonical form as indicated by the right pointing triangles along the rows. The rightmost two columns are then contracted and compressed to fixed bond dimension $D_I$ using the standard MPS compression algorithm, where the column is first converted into a canonical form (here, top canonical form) and then compressed by singular value decomposition in the reverse direction (leaving it in bottom canonical form). 
    The canonical form implies that the tensors satisfy an isometric condition (see diagram on the right); e.g. the right pointing arrow implies contraction of a tensor with its complex conjugate over the left, up, and down indices yields the identity matrix. 
    The procedure is repeated until all columns have been contracted. }
    \label{diag:IF_alg}
\end{figure*}

Alternatively, one may use standard matrix product state techniques to compress the intermediates that arise during the contraction to restrict bond dimensions $D_\rho$ or $D_I$ to some constant value \cite{Vidal_TE,Daley_TEBD,Schollwock_MPS}. 
\textcolor{black}{In this case, the cost of the algorithm is dominated by the cost of the MPS compression (which requires performing a series of singular value decompositions), which scales like $\mathcal{O}(K D_\rho^3 D_L^3 d_\rho)$ for the direct time evolution case (upward contraction by row), and $\mathcal{O}(N D_I^3 d_\rho^3 D_L )$ for the IF time evolution procedure (sideways contraction by column). }
In this paper, we use such an approximate transverse contraction scheme (compressing to a fixed bond dimension of the contraction intermediates) to compute the IF (Fig.~\ref{fig:spacetime}(b)) for systems with arbitrary baths. The algorithm is to iteratively contract and compress the columns from the edges of the bath inwards to the sites connected to the subsystem (Fig.~\ref{diag:IF_alg}). 
Assuming the subsystem is defined as the leftmost site, we start from the rightmost boundary column, and then the column is absorbed leftward to make a new boundary column, which is compressed using standard MPS compression to a pre-specified maximum bond dimension $D_I$. This is the standard % following the
"boundary contraction" algorithm of 2D tensor networks \cite{Murg_2DBH}.
\textcolor{black}{Key to the success of the algorithm and the quality of the compression is the choice in gauge (a redundant degree of freedom in all tensor networks). We choose the gauge as shown in Fig.~\ref{diag:IF_alg}.} 
\textcolor{black}{In the case where the bath extends both to the left and right of the site of interest, we compute the IFs corresponding to the left and right bath sites separately (see Appendix for details).}
This contraction scheme was implemented within the Quimb tensor network library~\cite{gray2018quimb}.

For purposes of comparison, we will also present reference dynamics generated by standard MPS time evolution (ie. contracting the space-time tensor network in the usual time direction) \cite{Vidal_TE, Vidal_2004_thermal, Ripoll_TE, Schollwock_MPS}. 
Because the underlying full system dynamics is governed by Hamiltonian evolution in the problems that we study, we have the option to apply $e^{-i H \epsilon}$ as a commutator to the square root of the density matrix ($[e^{-i H \epsilon}\rho^{1/2}][\rho^{1/2\dag} e^{i H \epsilon}]$) or via $e^{-i L\epsilon}$ directly. We refer to the former as Hilbert  time evolution (HTE) and the latter as Liouville time evolution (LTE). In HTE, the  compressed tensor network dynamics is carried out for the pseudowavefunction $\psi = e^{-i H \epsilon}\rho^{1/2}$ ~\cite{Verstraete_2004_thermal,Barthel_2009_thermalTE,Karrasch_2012_thermalTE}. HTE has the advantage that the compressed density matrix is always positive definite, although correlations between the bra and ket sides of the density matrix are less compressible. 
In existing literature, this is sometimes referred to  as purification-based time evolution. In the case that $\rho$ is a pure state, this method is equivalent to traditional MPS Hilbert space time evolution.

\section{Results}

\subsection{Spin Boson Model}
First, we consider the well-studied spin-boson model, in which a single spin is linearly coupled to a bath of non-interacting harmonic oscillators,
\begin{align}
    H_{SB} = \Delta S_X + \int\! d\omega\left[  S_Z (g(\omega) a_\omega + g^*(\omega) a^\dagger_\omega) + \omega a^\dagger_\omega a_\omega\right]
\end{align}
where $\Delta$ is the tunneling strength between the two subsystem states, and the system-bath coupling strength $g(\omega)$ is determined from the bath spectral density function $J(\omega)$ by 
\begin{align}
    |g(\omega)|^2 = \frac{1}{\pi} J(\omega)
\end{align}  
In the case of an Ohmic bath with exponential cut-off,
\begin{equation}
J(\omega) = \frac{\pi}{2} \alpha \omega e^{-\omega/\omega_c}
\end{equation}
where $\alpha$ is the Kondo parameter and $\omega_c$ is the cut-off frequency. Typically one computes the dynamics from a factorized initial state $ \kett{\rho_s(s_{t_0})} \kett{\rho_{b}(\beta)}$, where $\kett{\rho_b(\beta)}$ is the Gibbs thermal state of the isolated bath at finite temperature $\beta$. 
In this paper, we set $\Delta=1.0$, $\omega_c=7.5$, and $\beta=5.0$.

The spin-boson model exhibits a dynamical phase transition from thermalizing to localizing behavior at $\alpha=1.0 + \mathcal{O}(\Delta/\omega_c)$~\cite{Strathearn_TEMPO}, and is often cited as an example of physically relevant non-Markovian dynamics \cite{Breuer_book,Vega_nonmarkovian}. Because of the linear coupling and harmonic bath, the IF may be computed via an analytical expression. There already exist several methods of obtaining accurate dynamics for various bath coupling strengths and spectral densities \textcolor{black}{ \cite{Makri_1995, makri_2014_blip, makri_2020_smatpi, MCTDH_0K, Strathearn_TEMPO} }. We thus use this model as a benchmark to understand the properties of the influence functional, its compressibility, and the accuracy of the tensor network contraction approximation.

\begin{figure*}[t]
    \includegraphics[width=\linewidth,trim={0cm 0cm 0cm 0cm}]{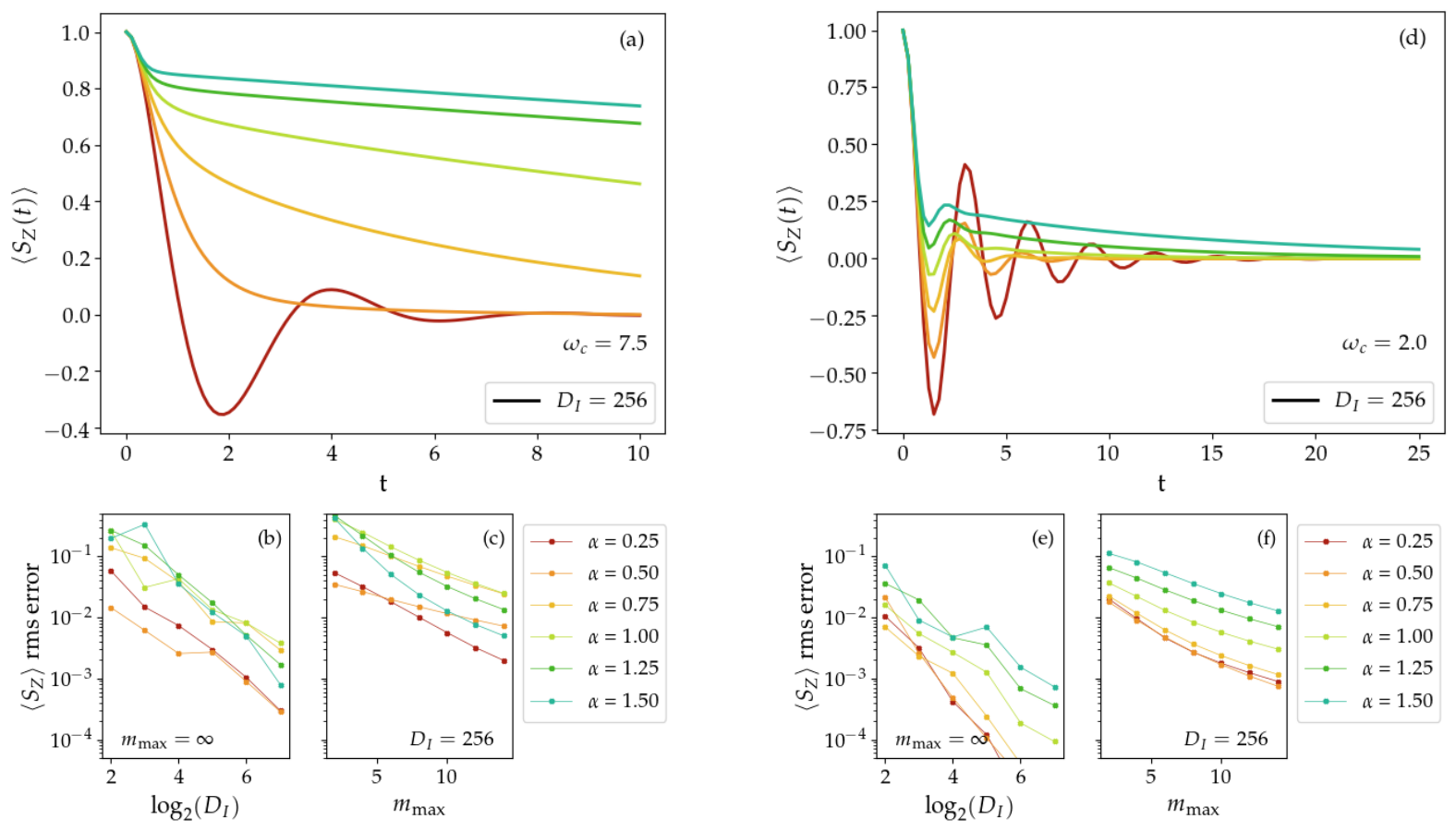}
    \caption{ \textcolor{black}{ (a, d) Dynamics obtained using the analytical IF with $m_{\max}=\infty$ and compressed to bond dimension $D_I=256$ for various coupling strengths $\alpha$. (b, e) R.m.s. error for the analytical IF with respect to $D_I$. (c, f) R.m.s error for capped IFs with respect to $m_{\max}$ computed using $D_I=256$. The errors are obtained using dynamics from the IF with no cap ($m_{\max}=\infty$) and $D_I=256$ as reference. % The error of the complete IF with respect to $D_I$ is independent of coupling strength. In contrast, the error for finite $m_{\max}$ is larger near the localization transition.
    These calculations are performed using $N=100$ time steps. The system parameters are $\Delta=1.0$, bath inverse temperature $\beta=5.0$, and an Ohmic bath spectral density with exponential cut-off $\omega_c = 7.5$ for (a-c) and $\omega_c=2.0$ for (d-f). }
    }
    \label{fig:sb_anl}
\end{figure*}

\subsubsection{Compressibility of Analytical IF}

We first investigate the compressibility of the analytical expression for the influence functional for the spin-boson model. 
Denoting the system basis $\kett{s} \equiv \ket{s^+}\bra{s^-}$ where $\ket{s^\pm} \in \{ \ket{1}, \ket{-1}\}$
and $\ket{1}, \ket{-1}$ are the eigenstates of the $S_Z$ operator, then the influence functional can be written as
\begin{align}
    I_{SB} &= \exp \left\{ - \sum_{k=1}^N \sum_{k'=1}^k (s_{t_k}^+ - s_{t_k}^-)(\eta_{kk'}s_{t_{k'}}^+ - \eta_{kk'}^* s_{t_{k'}}^- ) \right\} \nonumber \\
\label{eq:exactSB}
\end{align}
This explicitly shows the form of the influence functional as the Boltzmann weight of a complex spin Hamiltonian with the spins interacting along the time axis via the long-range pairwise "interaction" $\eta_{kk'}$. We can further factorize the weights into contributions for times $(t_1)$, $(t_1, t_2), \ldots, (t_1, t_{N})$, giving 
\begin{align}
    I_{SB} = & \prod_{k=1}^N I_0(s_{t_k}^\pm) \prod_{k=1}^{N-1} I_1(s_{t_k}^\pm,s_{t_{k+1}}^\pm) \, \dots  \nonumber \\
    & \prod_{k=1}^{N-m} I_{m} (s_{t_k}^\pm,s_{t_{k+m}}^\pm) \: \dots
    \prod_{k=1}^1 I_{N}(s_{t_k}^\pm,s_{t_{k+N-1}}^\pm)
\label{eq:factorexactSB}
\end{align}
where
\begin{equation}
    I_{m} = \exp \left\{ -(s_{t_{k+m}}^+ - s_{t_{k+m}}^- ) (\eta_{k+m,k} s_{t_k}^+ - \eta_{k+m,k} s_{t_k}^-)  \right\}
\end{equation}
The $\eta_{kk'}$ interaction terms can be derived from the spectral density of the bosonic bath. Expressions for $\eta_{kk'}$ are given in Eq.~(12) of Ref.~\onlinecite{Makri_1995}, where they use $\Delta t$ instead of $\epsilon$ to denote the timestep.

In QuAPI, one considers interactions $\eta_{kk'}$ for times $k$ and $k'$ within $m_{\text{max}}$ of each other, where $m_{\text{max}}$ is treated as a convergence parameter \cite{Makri_1995}. Then,
one can evaluate the influence functional (or its effect on the dynamics~\cite{makri_2020_smatpi, makri_2020_memorylength}) with a computational cost exponential in $m_\text{max}$.
Since QuAPI often converges rapidly with $m_\text{max}$, %Under similar assumptions,
one also expects the matrix product state representation of the IF (Eq.~\eqref{eq:IF_MPS}) to be compressible to small bond dimension. One way to verify this would be to construct the large influence functional object as an exact tensor, and then compress it into a matrix product state. Because of the exponential storage of the tensor with time, this is possible only for a small number of time points $N$. Alternatively, one could build the influence functional iteratively (i.e. piece by piece in Eq.~\eqref{eq:factorexactSB}) and compress at each step. This is the idea behind TEMPO and related methods~\cite{Strathearn_TEMPO, Jorgensen_TEMPO} which exploit the compressibility of the augmented density matrix, the influence functional applied to the subsystem density matrix, i.e.
\begin{align}
A(s_{t_1},\ldots,s_{t_{m_{\max}}}) = \sum_{s_{t_0}} I(s_{t_0}, \ldots, s_{t_{m_{\max}}}) \rho_s(s_{t_0}) \,\, .
\end{align}
Note that because $I$ is composed of commuting pieces there are many possible decompositions and thus sequences of iterative constructions.

To verify the compressibility of the IF itself, we approximate $I_{SB}$ 
\textcolor{black}{with no memory cutoff ($m_\text{max}=\infty)$}
as an MPS of bond dimension $D_I$ using an iterative scheme (see Appendix) and determine the error in the resulting on-site dynamics, using the \textcolor{black}{$D_I=256$}  % $D_I=128$
result as reference. Here and throughout the paper, the error is computed as the r.m.s. deviation of the dynamics of an observable with respect to some reference over the time interval of the plot. 
\textcolor{black}{The $D_I=256$ results are used as the reference because most of the $D_I=128$ dynamics calculations are already converged to within an r.m.s. error of 0.001 with respect to it.}

Fig. \ref{fig:sb_anl} compares convergence with respect to $D_I$ for infinite $m_{\text{max}}$, as well as convergence with respect to $m_{\text{max}}$ for fixed \textcolor{black}{$D_I=256$}.
% We find that the analytical influence functional is relatively compressible for the parameter sets considered here. 
\textcolor{black}{As expected, the influence functional generally becomes less compressible with increased Kondo parameter $\alpha$.} 
\textcolor{black}{However, while convergence of the IF with respect to $m_\text{max}$ and $D_I$ seem comparable for baths with larger cut-off frequencies ($\omega_c=7.5$), the IF converges more quickly with respect to $D_I$ for smaller $\omega_c$.}
% \textcolor{red}{However, while the convergence of the IF with respect to $m_\text{max}$ is slower for $\omega_c$=2.0 compared to the $\omega_c=7.5$ case, the IF convergence with respect to $D_I$ is relatively insensitive to the different values. } % Thus, while similar, compressed tensor network representations of the IF are not equivalent to QUAPI methods
% Even at a small bond dimension of $D_I=16$, the correct behavior of the dynamics in both the thermalization and localization regimes is captured. % \cite{MCTDH_0K, Strathearn_TEMPO}. 
% \textcolor{black}{Compared to results for small $m_{\max}$,} the MPS algorithm yields slightly more accurate dynamics near the localization transition. 
The key advantage of using a compressed matrix representation, as opposed to truncating $\eta_{k,k+m}$ at some $m_{\max}$ as in QUAPI, is that this does not eliminate the effects of long-range memory \cite{Strathearn_TEMPO}.
Overall, this result confirms that for certain spectral densities, the influence functional can be efficiently written as a low-rank MPS, \textcolor{black}{consistent with the findings from TEMPO\cite{Strathearn_TEMPO,Jorgensen_TEMPO}}.

\subsubsection{Finite Size Harmonic Bath}

\textcolor{black}{We now use the spin-boson model to examine if the IF can be constructed efficiently from our tensor network contraction scheme. To do so, we consider a finite size harmonic bath with $K$ sites, where the bosons are capped to some finite number of states. 
For small baths and small boson cap, the transverse contraction can be performed without compression, allowing a numerical test of the compression procedure.}   
\textcolor{black}{The analytical IF, assuming  bosons with infinite boson cap, can also be computed.}
The bath is characterized by the discretized spectral density, 
\[ J_D(\omega) = \sum_j \frac{J(\omega)}{\rho(\omega)} \delta(\omega-\omega_j) \] We use a linear discretization of the bath sites, such that the bath density is $\rho(\omega) = K/\omega_m$, where $\omega_m$ is the maximum boson frequency used. Here, we set $\omega_m = 10$. 

\begin{figure}[t]
    \centering
    \includegraphics[width=\linewidth,trim={0cm 0cm 0cm 0cm},clip]
    {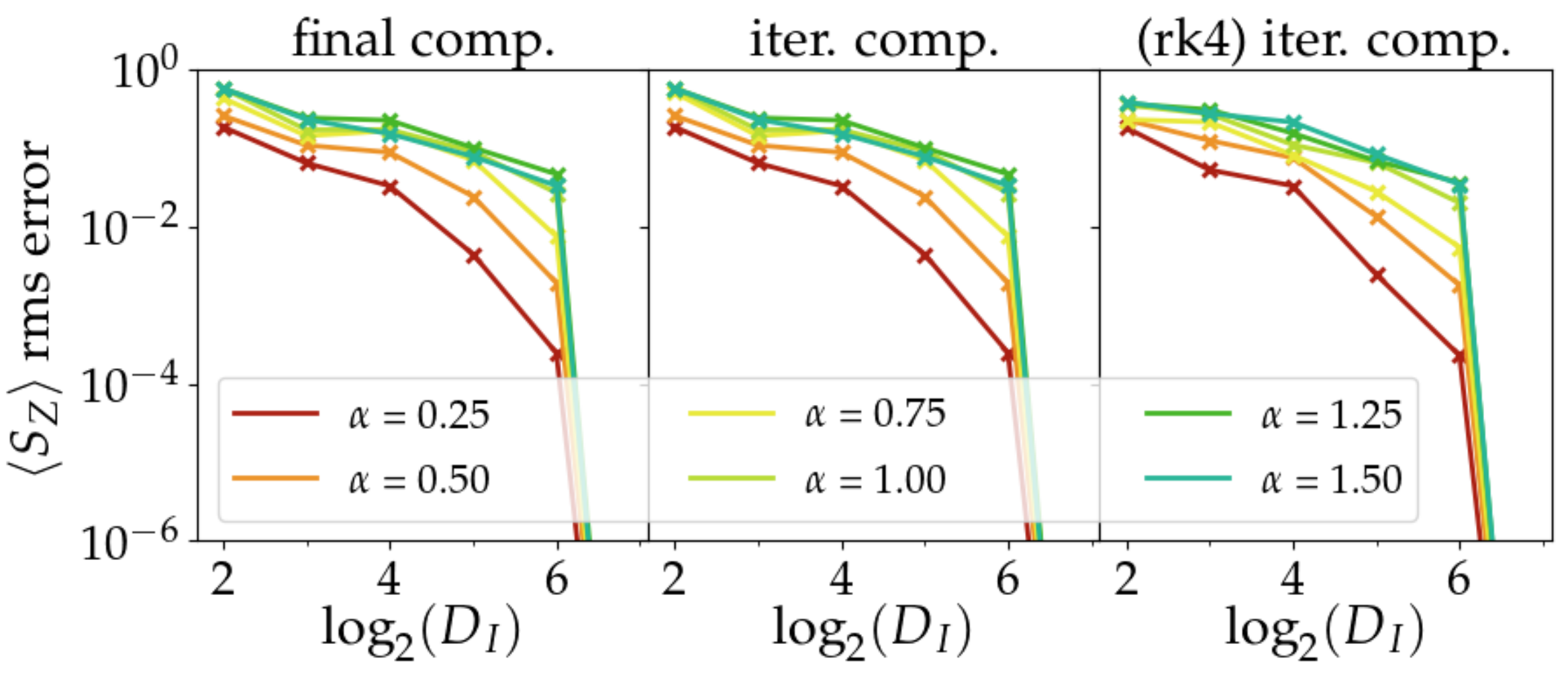}
    \caption{Comparisons of average error in $\avg{S_Z(t)}$ with respect to reference direct time evolution results for the IF where the IF is (left) constructed using exact time evolution, contracted exactly, and finally compressed to bond dimension $D_I$ at the end, and where the IF is constructed with (middle) exact time evolution and (right) RK4 time evolution but iteratively contracted and compressed using the transverse compression scheme. In these calculations, we use $\Delta=1.0$, bath inverse temperature $\beta=5.0$, cut-off frequency $\omega_c=7.5$, and assume a discrete Ohmic bath with 2 modes at $\omega=[5.0,10.]$. The bosonic bath sites are approximated to have only a maximum boson number of 2. Time evolution is performed using $N=100$ time steps with a time step of $\epsilon=0.05$. The plots show that for small bath sizes, the error in the iterative compression scheme is dominated by the lack of compressibility of the final IF. }
    \label{fig:sb_small}
\end{figure}

First, we consider a system with only 2 bath modes \textcolor{black}{(in the number basis)} each with a maximum boson number of 2, for $N=100$ time steps of size $\epsilon = 0.05$. For this small system, we use the exact time evolution operator of $L_{bs}$ and compute the  IF by exact transverse contraction, applying compression only to the final IF object. Fig.~\ref{fig:sb_small}(a) shows the error of the exact IF compressed to bond dimension $D_I$. As expected, the IF is much less compressible than with the continuous bath density in the last section, due to the small bath size. The error decreases only slightly until it drops suddenly once the bond dimension is large enough to capture the IF exactly, and further, the compression error increases with $\alpha$.
We then perform the same analysis but for IFs computed by transverse contraction with compression, for both exact and RK4 time evolution. As seen in Fig. ~\ref{fig:sb_small}, the errors are comparable to those obtained when compressing the final exact IF. This indicates that for this small problem, there is little additional error added by the iterative contraction, and that the time-step error is negligible: the error is dominated by the compressibility of the final IF itself (which is low when the bath size is small).

\begin{figure}[t]
    \includegraphics[width=\linewidth,trim={0cm 0cm 0cm 0cm}]{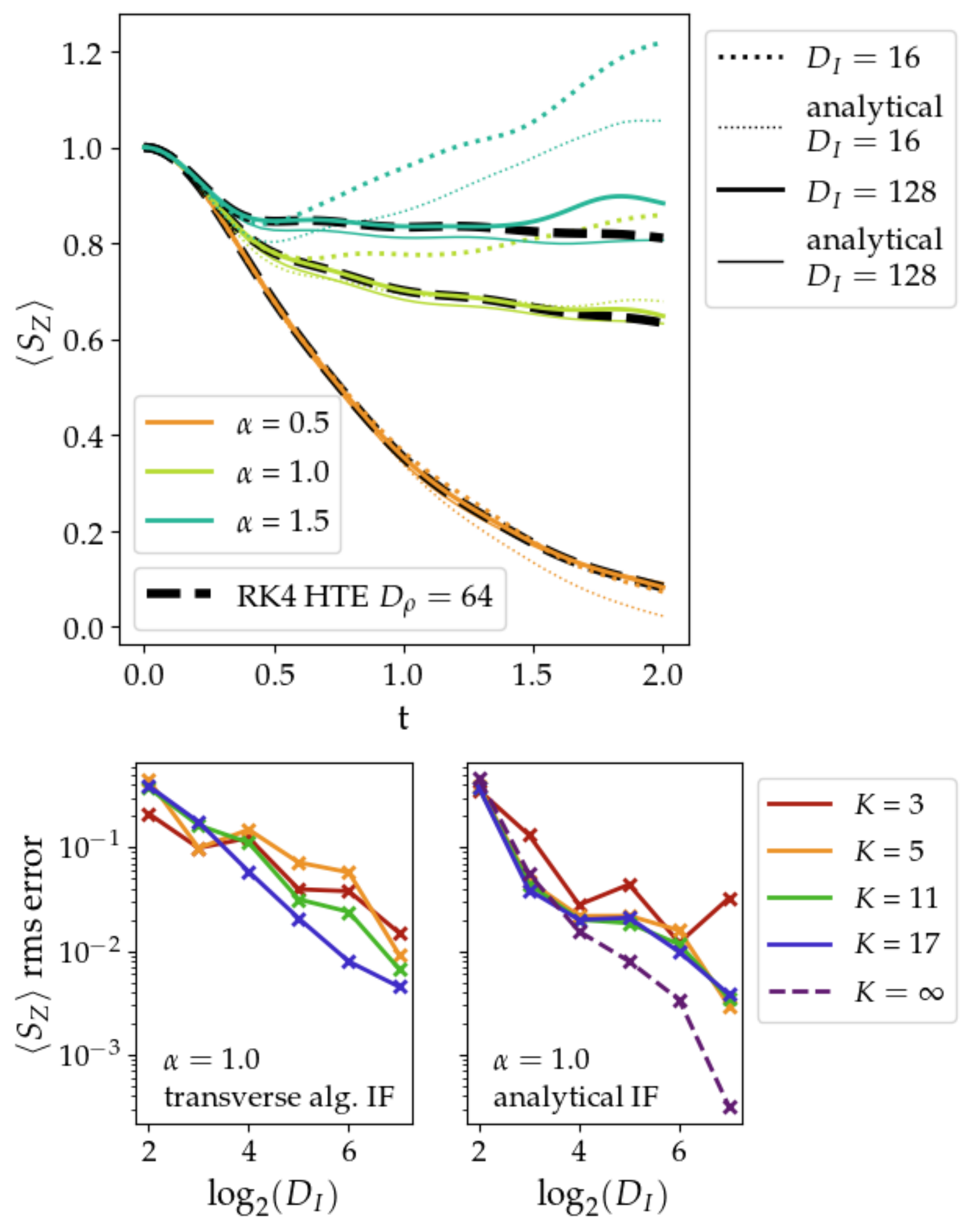}
    \caption{(top) Expectation values $\avg{S_Z(t)}$ obtained from the analytical IF (thinner colored lines) and IF from transverse contraction (thicker colored lines) for bath size $K = 11$ with $\alpha=$ 0.5, 1.0, and 1.5. The thick dashed line corresponds to reference dynamics from direct time evolution of the density matrix. The transverse contraction scheme introduces additional error with respect to the analytical result, which increases with $\alpha$. (bottom) Time-averaged error in IF dynamics with respect to $D_I=128$ results, obtained using (left) the transverse contraction scheme and (right) the analytical IF for the finite bath of size $K$, respectively. The $K=\infty$ bath size corresponds to a continuous bath.}
    \label{fig:sb_dynamics}
\end{figure}

\begin{figure*}[t!]
    \centering
    \includegraphics[width=\linewidth,trim={-0.5cm 0cm 0.5cm 0cm}]
        {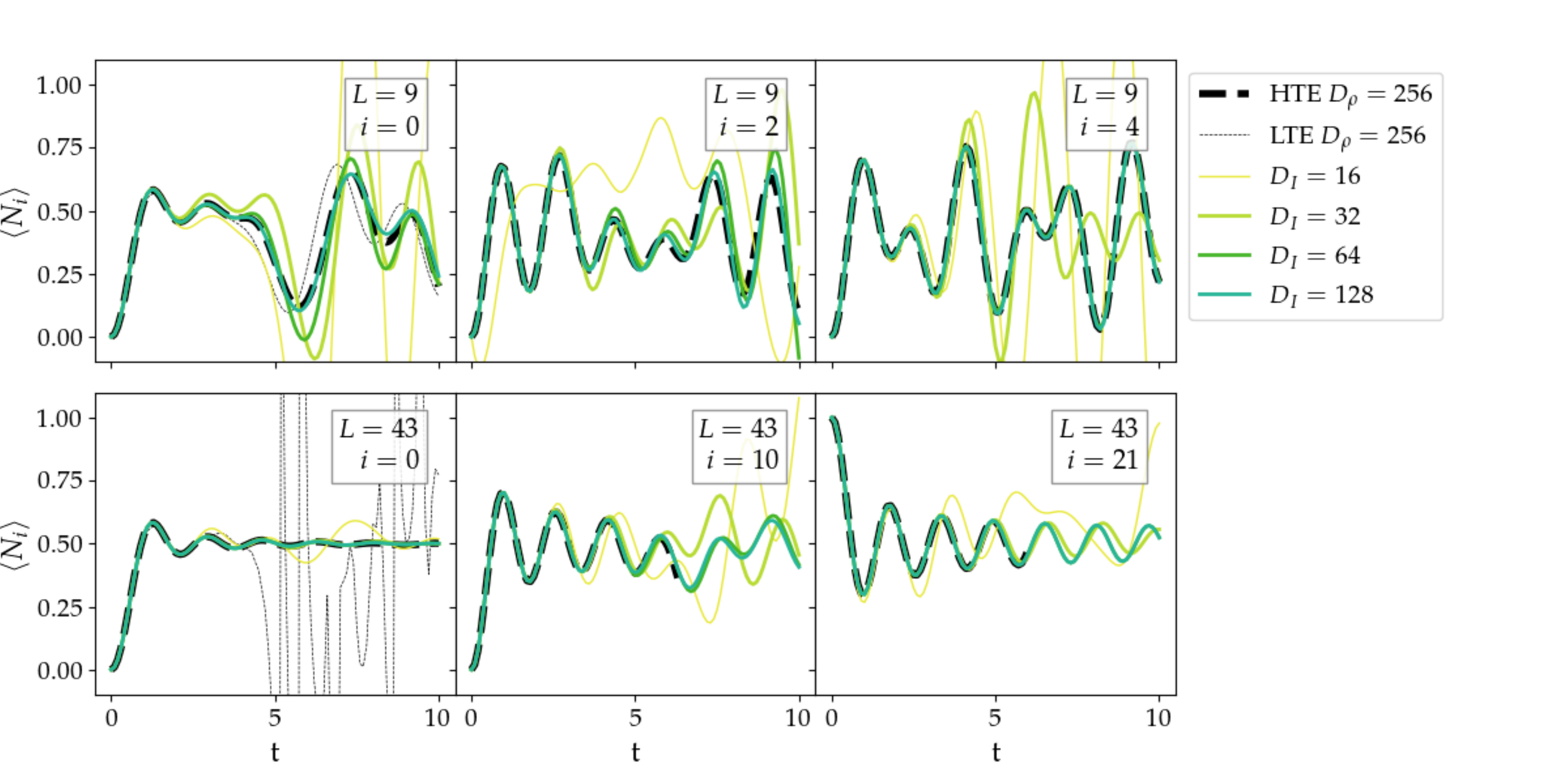}
    \caption{Expectation values $\avg{N_i(t)}$ for the hard-core boson model with $U=0$ for system sizes (top) $L=9$ and (bottom) $L=43$ for sites $i=\{0,L/2,L/4\}$ obtained using the iterative contraction scheme with bond dimension $D_I$ as labeled. 
    The thick dashed line is the dynamics obtained by direct Hilbert-space time evolution, compressed to bond dimension $D_\rho=256$, shown only up to times for which the results are converged.
    The thin dashed line, shown only in the left-most plots but with similar behavior for all, is the dynamics obtained by direct time evolution of the full density matrix. Unphysical behavior suggests loss of positivity of the density matrix. For these calculations, the initial state is a pure product state with alternating spins, $\ket{0,1,0,1,...}$. Time evolution is performed using Trotter steps with a $N=100$ time steps of $\Delta t=0.1$.
    Consistent with earlier observations, larger $D_I$ is needed to accurately capture the IF for smaller bath sizes. However, as shown in the $L=43, i=21$ simulation, the IF of comparable bond dimension can simulate dynamics for longer times than direct HTE. }
    \label{fig:bh_dynamics}
\end{figure*}

Next we investigate systems with larger bath sizes in Fig.~\ref{fig:sb_dynamics}(a). We first examine the time-dynamics of the analytical IF (i.e. without any boson cap, and without transverse contraction) for a discretized spectral density with 11 bath sites, as well as the IF computed by transverse contraction, using a boson cap of 2. 
Because the system size is so small, the exact reference dynamics for a boson cap of 2 can be generated by direct MPS time evolution (here we use $D_\rho=64$ and Hamiltonian time evolution). From this comparison we observe two things. First, compared to using the continuous bath density, the error of the analytical IF dynamics is increased, although it is still somewhat compressible. 
For the same $D_I$, the errors using transverse contraction are larger, suggesting that at intermediate points in the transverse contraction, there is more time-like entanglement than in the final IF itself. In Fig.~\ref{fig:sb_dynamics}(b) we show the time-averaged error of the IF dynamics as a function of the number of bath sites. We see that this error decreases as the number of bath sites increases, both for the analytical IF and the transverse contraction. This is consistent with the idea that smoother bath densities are more "compressible".

\subsection{1D Hard-core Boson Model}

\textcolor{black}{We next study dynamics of a 1D hard-core boson (HCB) lattice model, in which each lattice site is either unoccupied ($\ket{0}$) or occupied ($\ket{1}$) by a single bosonic particle. We consider the Hamiltonian}
\begin{equation}
    H_{HCB} = \sum_j \left[ -J(a_j^\dagger a_{j+1} + h.c.) + U n_j n_{j+1} + \frac{K}{2} n_j j^2 \right]
\end{equation}
where \textcolor{black}{$a_j^\dagger = \ket{1}\bra{0}$, $a_j = \ket{0}\bra{1}$ are hard-core boson creation and annihilation operators at the $j^\text{th}$ lattice site, and $n_j = a^\dagger_j a_j$ is the number operator.} 
\textcolor{black}{This Hamiltonian is intended to mimic the Bose-Hubbard model dynamics often simulated by cold-atom experiments, which was shown to be difficult to compute using direct time evolution methods \cite{Trotzky_BH}, where the on-site interaction term is replaced by a nearest-neighbor interaction term. We also include a harmonic potential term to emulate a cold atom trap.}
We assume a pure initial state $\ket{0,1,0,...}$ such that there is one particle at every other lattice site, and set the parameters $J=1$ and $K=10^{-2}$ while varying $U$. For non-zero interaction term $U$, there is no analytical form for the IF.

\begin{figure*}[t]
    \includegraphics[width=\linewidth,trim={0cm 0cm 0cm 0cm},clip]{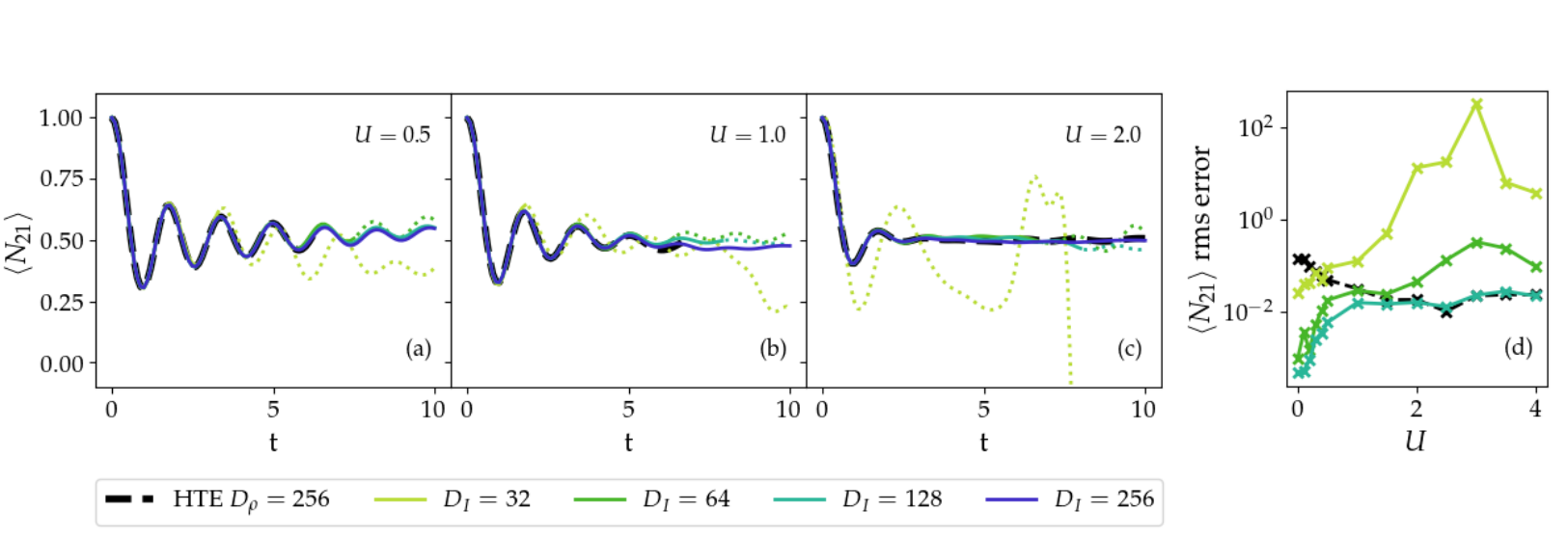}
    \caption{(a,b,c) Expectation values of the site-occupancy $\avg{N_i(t)}$  for the hard-core boson model of length $L=43$ at lattice site $i=21$ for different coupling strengths $U=$ 0.5, 1.0, and 2.0. Lines become dotted after divergence of >0.03 with respect to the $D_I=256$ IF results. (d) The r.m.s. errors with respect to the $D_I=256$ IF results as a function of $U$.
    Simulation parameters are otherwise the same as in Fig. \ref{fig:bh_dynamics}. Compared to the $U=0$ case, a larger bond dimension is needed, particularly at around $U=3.0$ where the r.m.s. error peaks. In contrast, the HTE dynamics converge more quickly with increasing $U$.}
    \label{fig:bh_dynamics_U}
\end{figure*}

We compute the dynamics of $\avg{N_i(t)}$ at lattice sites $i=\{0, L/4, L/2\}$, where $L$ is the length of the 1-D chain. For longer chains, the rapid growth of entanglement means that direct MPS time evolution (either using Hamiltonian evolution, denoted HTE or Liouvillian evolution, denoted LTE) with a finite $D_\rho$ can only obtain converged dynamics up to a finite time. We consider two chain lengths: $L=9$ where converged HTE MPS dynamics can be used as a reference, and $L=43$,
{where the HTE MPS dynamics appear to be not fully converged (no longer within 0.03 of $D_\rho=512$ results) for the full simulated time.}
To obtain the dynamics using the influence functional method, we partition the lattice such that site $i$ is the subsystem of interest and the remaining sites are the bath. 

The $U=0$ dynamics for $L=9$ and $L=43$ is shown in Fig.~\ref{fig:bh_dynamics}. Direct MPS LTE shows unphysical behaviour for large system sizes, presumably because of the loss of positivity at some point in the dynamics. In contrast, the dynamics obtained using the iteratively contracted IF are more stable, highlighting the innate compressibilty of time evolution tensor network along the time axis as opposed to the spatial axis. 
For $L=9$, the IF dynamics only appears to begin to converge by $D_I=128$ with respect to the (exact) HTE dynamics, having less than 0.03 r.m.s error in $\avg{N_i(t)}$ over the simulated time interval and deviations within 0.08. Thus, there appears to be no significant advantage to using the the IF method over direct HTE for small system sizes. Conversely, for the $L=43$ system, the IF dynamics are converged with respect to $D_I=256$ results by $D_I=64$, with less than 0.02 r.m.s error and a maximum deviation of 0.04 for $i=10$ and less than 0.001 r.m.s error and a maximum deviation of 0.004 for $i=21$.
Note that the $i=10$ dynamics converge more slowly because effectively the site is coupled to two separate baths, one of which is small. However, the IF method still outperforms direct HTE.

For $U>0$, as shown in Fig.~\ref{fig:bh_dynamics_U}, the $D_I=64$ results appear less converged than the non-interacting case, but the $D_I=128$ results are converged (r.m.s. errors of the $D_I=128$ observable dynamics with respect to the $D_I=256$ results are less than 0.03), for times longer than that accessible by direct time evolution. This shows that the IF-based dynamics can produce the correct oscillatory behaviour of the density as a function of time, which is not captured by the direct MPS time evolution despite using a larger bond dimension (this difficulty with the long time oscillatory tail has previously been noted in other cold atom simulations~\cite{Trotzky_BH}). However, while the IF method notably outperforms direct HTE at small $U$, the two methods become comparable at larger $U \approx 1.5$ once the oscillatory tail is sufficiently dampened.

\subsection{Entanglement Spectrum}

\begin{figure}[t]
    \includegraphics[width=\linewidth,trim={0cm 0cm 0cm 0cm},clip]{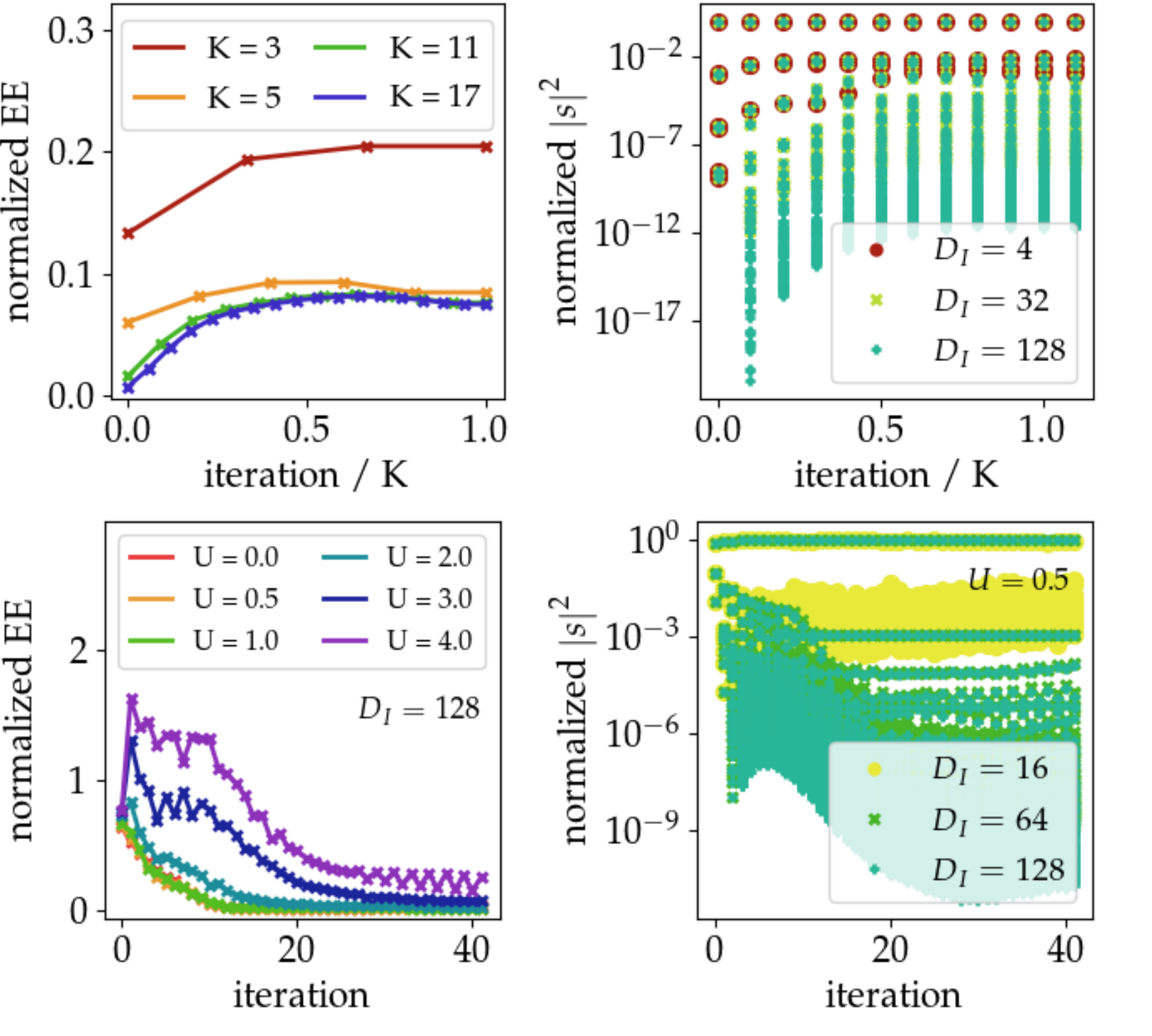}
    \caption{(Top left) Entanglement entropy and (top right) spectrum of normalized singular values at the middle of the "bath" boundary column MPS after each contraction and compression iteration for the SB model ($\alpha=1.0$). As bath size increases, the EE of the IF decreases, converging to some finite value. (Bottom left) Entanglement entropy of the hard-core boson model with $L=43$ for the "bath" boundary MPS with contractions starting from the right edge, plotted for different values of $U$. The decrease in EE with respect to iteration shows that EE decreases with system size. (Bottom right) Normalized singular values for the hardcore-boson model with $L=43$ and $U=0.5$. Surprisingly, for insufficient $D_I$, the singular values take on large and incorrect values, yielding an artificially large EE.}
    \label{fig:EE}
\end{figure}

The accuracy of the transverse contraction scheme depends on the entanglement in the time-like direction. Recall that our contraction algorithm starts with the farthest column (an MPS), and at each iteration another column is contracted into this boundary. Thus, as the iteration number increases, the boundary column represents more of the bath. For both the spin-boson model and HCB model, we measure the singular values at the middle of the boundary "bath" MPS during each step of the iterative contraction scheme. The entanglement entropy (EE) and spectrum of the singular values (normalized so that $\sum_i |s_i|^2 = 1$) are plotted in Fig.~\ref{fig:EE}. For the spin-boson model, only results for $\alpha=1.0$ are shown; the only notable difference for other $\alpha$ is that the EE increases with $\alpha$.
Consistent with the observations in our simulations above, the EE decreases as one increases bath size. For the SB model, the EE decreases with increased number of bath sites in the discretization until convergence. For the HCB model, {\it only if} sufficiently large enough bond dimension is used does the EE decrease with increasing iteration number. 
Otherwise, the EE stays at a large value throughout the contraction scheme and the gap between the dominant and non-dominant singular values decreases; this makes the EE of the smaller $D_I$ approximation larger than that of the larger $D_I$ approximation. Overall, this suggests that the final compressibility of the IF emerges from the cancellation of many different correlations as one iteratively contracts out the bath.

\section{Conclusions}

In this work, similar to some other recent contributions~\cite{Lerose_IFmatrix,Cygorek_IFmpo}, we have used the representation of the influence functional within the tensor network language, motivated by the limitations of modeling spatial entanglement growth in quantum dynamics. We have discussed a transverse tensor network contraction algorithm that allows us to compute the influence functional in cases where the analytical form is not known. We have applied this algorithm to study both the canonical spin-boson model as well as an interacting hard-core boson chain where the bath is not quadratic (i.e. interacting).  
We find that the compressibility of the influence functional is controlled by several factors, principally the size of the bath, as well as the nature of the interactions. In addition, although the time-like correlations may ultimately be short-ranged in the final influence functional, during the transverse tensor network contraction to construct it, it is possible to proceed through intermediate quantities with larger time-like entanglement. This suggests a complicated picture where time-like correlations first accumulate as the bath is integrated out before finally cancelling in the influence functional itself. In the regimes where the influence functional and all intermediate quantities are compressible, as in some interaction regimes in the interacting hardcore boson model we have studied, it is possible to outperform conventional tensor network time evolution methods at longer times.

There are many possible directions for further investigation. For example, there are natural extensions to higher-dimensional interacting problems and fermionic systems, as well as more complicated correlation functions. Also, a better theoretical understanding of how correlations grow and cancel out in the transverse contraction scheme may lead to a deeper understanding of the generation of memory in quantum dynamics, the master equation formalism~\cite{GQME_2003}, improved contraction schemes, and ultimately new algorithms to carry out longer time dynamical simulations. 

\section{Acknowledgments}
GKC was supported by the Center for
Molecular Magnetic Quantum Materials, an Energy Frontier Research Center funded by the U.S. Department of Energy, Office of Science, Basic Energy Sciences under Award No. DE-SC0019330. EY was primarily supported by the Google PhD fellowship program with supplemental salary support from the Center for Molecular Magnetic Quantum Materials. Support for industrial mentorship for EY was provided via the QISE-NET program, funded via NSF award DMR-1747426.

\section{Data and Code Availability}
Data and code is available upon request.

\appendix

\section{Transverse Contraction Procedure}

\begin{figure}
    \centering
    \includegraphics[width=\linewidth,trim={0cm 0cm 0cm 0cm}]{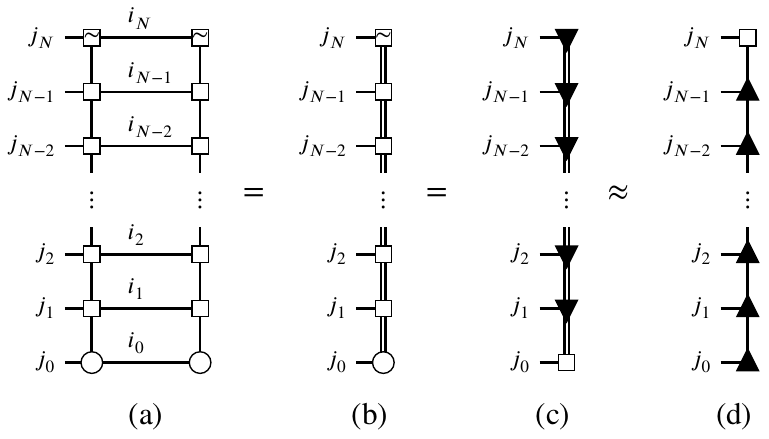}
    \caption{Tensor network diagrams depicting the contraction and compression of an MPO-MPS product. (a) The MPO and MPS columns to be contracted together. (b) The resulting MPS after exactly contracting the network along the horizontal bonds indexed by $\{ i \}$. We use a double line to represent the merging of the two original vertical lines, as in Eq.~\eqref{eq:col_contract}. (c) The new MPS put into right canonical form. (d) The MPS is compressed to a smaller bond dimension starting from the left-most tensor and moving to the right, and the final MPS is now in left canonical form. The double line is reduced to a single line to signify the size reduction of the large vertical bonds down to the desired bond dimension. 
    }
    \label{fig:si_contract}
\end{figure}

We describe a single step of the transverse contraction procedure, which in essence is performing an MPO-MPS contraction and then compressing the resulting MPS to the desired bond dimension $D$, described in detail in Ref.~\onlinecite{Schollwock_MPS}. We assume that all tensors have been prepared in the appropriate gauge (discussed later) as shown in Fig.~\ref{diag:IF_alg}. This algorithm is also the standard boundary contraction algorithm for 2D tensor networks, as described in Ref.~\onlinecite{Murg_2DBH}.

For simplicity, consider the influence functional tensor network for $K$ bath sites, as discussed in the main text. We start with the farthest boundary column MPS (the column for the $K^{\text{th}}$ bath site) which is given by 
% \begin{align}
%     P_K & (i_{K-1}^0,...,i_{K-1}^N) \\[0.25cm]
%     = & \sum_{\{ b^K \} } C_{i_{K-1}^{t_0}}^{(K)} (b^K_0) \, M^{(K)}_{i_{K-1}^1}(b_{t_1}^K,b_{t_0}^K)  \ldots \nonumber \\ 
%     & \hspace{1cm} \ldots M^{(K)}_{i_{K-1}^{N-1}}(b_{t_{N-1}}^K,b_{t_{N-2}}^K) \, \tilde{M}^{(K)}_{i_{K-1}^{N}}(b^K_{t_{N-1}}) \nonumber \\[0.25cm]
%     \simeq & \sum_{\{ b^K\} } A^{(0)}_{b^K_0}({i_{K-1}^0}) A^{(1)}_{b^K_0,b^K_1}(i_{K-1}^{1}) \ldots A^{(N)}_{b^K_{N-1}}(i_{K-1}^N) \nonumber
% \end{align}
% where $C^{(K)}$ and $M^{(K)}$ are the right-most tensors of the density matrix MPS (Eq.~\ref{eq:rho_mps}) and Liouville evolution operator (Eq.~\ref{eq:expLT_mpo}), respectively, and $\tilde{M}^{(K)}$ denotes the Liouville evolution operator with traced out bath degrees of freedom, $\text{Tr}_{b^K_{T}}(M^{(K)})$.
%

\begin{align}
    P_K & (i_0,...,i_N) %\nonumber %\\[0.25cm]
    = %& 
    \sum_{\{ a \} } A^{(0)}_{a_0}({i_0}) A^{(1)}_{a_0,a_1}(i_1) \ldots A^{(N)}_{a_{N-1}}(i_N)
    \label{eq:col_mps}
\end{align}
with
\begin{subequations}
\begin{align}
    & A^{(0)}_{a}({i}) = C^{(K)}_{i} (a) \\[0.15cm]
    & A^{(n)}_{a,a'}(i) = M^{(K)}_{i} (a',a)  \qquad \text{for} \,\, n = 1, ..., N-1 \\[0.15cm]
    & A^{(N)}_{a}(i) = \tilde{M}^{(K)}_{i} (a)
\end{align}
\end{subequations}
%\begin{subequations}
%\begin{align}
%    & A^{(0)}_{b^K_0}({i_{K-1}^0}) = C^{(K)}_{i_{K-1}} (b^K) \\[0.15cm]
%    & A^{(n)}_{b^K_{n-1},b^K_n}(i_{K-1}^{n}) = M^{(K)}_{i_{K-1}} (b_t^K,b_{t'}^K) \nonumber \\
%    & \hspace{3cm} \qquad \text{for} \,\, n = 1, ..., N-1 \\[0.15cm]
%    & A^{(N)}_{b^K_{N-1},b^K_N}(i_{K-1}^{N}) = \tilde{M}^{(K)}_{i_{K-1}} %(b_{t_{N-1}}^K)
%\end{align}
%\end{subequations}
where $C^{(K)}$ and $M^{(K)}$ are the right-most tensors of the density matrix MPS (Eq.~\eqref{eq:rho_mps}) and Liouville evolution operator (Eq.~\eqref{eq:expLT_mpo}), respectively, and $\tilde{M}^{(K)}$ denotes the Liouville evolution operator with traced out bath degrees of freedom. %, $\text{Tr}_{b}(M^{(K)}(b))$.

The next column can be interpreted as an MPO, given by 
% \begin{align}
%     & P_{K-1} (i_{K-2}^{0},...,i_{K-2}^{N}; i_{K-1}^{0},...,i_{K-1}^{N}) \\[0.25cm]
%     & \hspace{0.05cm} = \sum_{\{ b^{K-1} \} } C_{i_{K-2}^{0},i_{K-1}^{0}}^{(K-1)} (b^{K-1}_{t_0}) \,\, M^{(K-1)}_{i_{K-2}^{1},i_{K-1}^{1}}(b_{t_1}^{K-1},b_{t_0}^{K-1}) \ldots \nonumber \\
%     & \hspace{1.25cm} \ldots M^{(K-1)}_{i_{K-2}^{N-1},i_{K-1}^{N-1}}(b_{t_{N-1}}^{K-1},b_{t_{N-2}}^{K-1}) \,\, \tilde{M}^{(K-1)}_{i_{K-2}^N,i_{K-1}^N}(b_{t_{N-1}}^{K-1}) \nonumber \\[0.25cm]
%     & \hspace{0.05cm} \simeq \sum_{\{ b^{K-1} \} } B^{(0)}_{b_0^{K-1}}({i^0_{K-2},i^0_{K-1}}) \, B^{(1)}_{b_0^{K-1},b_1^{K-1}}(i^1_{K-2},i^1_{K-1}) \ldots \nonumber \\
%     & \hspace{4.5cm} \ldots B^{(N)}_{b_{N-1}^{K-1}} (i^N_{K-2}, i^N_{K-1}) \nonumber 
% \end{align}
\begin{align}
    & P_{K-1} (j_{0},...,j_{N}; i_{0},...,i_{N})  \nonumber \\[0.25cm]
    & \hspace{0.05cm} = \sum_{\{ b \} } B^{(0)}_{b_0}({j_0,i_0}) \, B^{(1)}_{b_0,b_1}(j_1,i_1) \ldots  B^{(N)}_{b_{N-1}} (j_N, i_N) \,\, .
    \label{eq:col_mpo}
\end{align}
with
\begin{subequations}
\begin{align}
    & B^{(0)}_{b}(j, i) = C^{(K-1)}_{j, i} (b) \\[0.15cm]
    & B^{(n)}_{b,b'}(j,i) = M^{(K-1)}_{j,i} (b',b) % \nonumber \\
    % & \hspace{3cm} 
    \qquad \text{for} \,\, n = 1, ..., N-1 \\[0.15cm]
    & B^{(N)}_{b} (j, i) = \tilde{M}^{(K-1)}_{j,i} (b)
\end{align}
\end{subequations}
Contracting the two columns yields a new MPS,
\begin{align}
    & \sum_{ \{i_{K-1}\} } P_{K-1}(j_0,...,j_N; i_{0},...,i_{N}) \,\, \times \,\, % \\[-0.1cm]
    % & \hspace{4cm} 
    P_{K}(i_{0},...,i_{N}) \nonumber 
    \\ %[0.35cm]
    & \hspace{1.0cm}= \sum_{\{ c \} } F^{(0)}_{c_0}({i_0}) F^{(1)}_{c_0,c_1}(i_{1}) \ldots F^{(N)}_{c_{N-1}}(i_N) 
\end{align}
where the sum over $c$ indicates sums over all $(a,b)$, and
\begin{align}
    & F^{(n)}_{c,c'}(i) = % \nonumber \\[0.25cm]
    % & \hspace{0.5cm}
    \sum_{i} B^{(n)}_{b,b'} (j,i) \, A^{(n)}_{a,a'} (i)
    \label{eq:col_contract}
\end{align}
for all $n$ indexing the tensors in the MPO and MPS. Thus, if the MPS was of bond dimension $D_K$ and the MPO was of bond dimension $D_{K-1}$, the new MPS has a bond dimension of $D_K D_{K-1}$. 
We then must compress this new MPS back to one of a lower bond dimension $D$ (again, see Ref.~\onlinecite{Schollwock_MPS} for a more in depth discussion). To do so, we first put the MPS in right or left canonical form. 

In left canonical form, all tensors in the MPS (except for the rightmost one) satisfies
\begin{align}
    \sum_{l,u} L_{l,r}(u) \,\,\, & L^{*}_{l,r'}(u) = \delta_{r,r'} % \mathbb{I}\,(r,r') 
    \label{eq:left_mps}
\end{align}
where $\delta$ is the Kronecker delta.
The tensor network diagram depicting Eq.~\eqref{eq:left_mps} is
\begin{align*}
    \begin{tikzpicture}[baseline,scale=0.5]
        % \draw[thick,rounded corners] (-0.5,-0.125) -- (-0.5,2.625) -- (-1.5,2.625) -- (-1.5,-0.125) -- cycle;
        \draw[thick,rounded corners] (-0.5,0.5) -- (-0.5,2.0);
        \draw[thick,rounded corners] (0.25,0.5)--(-1.25,0.5)--(-1.25,2.0)--(0.25,2.0);
        \node at (-0.20,1.25) {\scriptsize $u$};
        \node at (-1.0,1.25) {\scriptsize $l$};
        \draw[fill=black] (-0.65,2.2)--(-0.25,2.0)--(-0.65,1.8)--cycle;
        \draw[fill=black] (-0.65,0.7)--(-0.25,0.5)--(-0.65,0.3)--cycle;
        \node[rotate=0] at (1.0,1.25) {=};
        \draw[thick,rounded corners] (3.75,0.5)--(2.5,0.5)--(2.5,2.0)--(3.75,2.0);
        \draw[fill=white] (2.0,0.75) rectangle (3.0,1.75);
        \node at (2.5,1.25) {$\mathbb{I}$};
    \end{tikzpicture}
\end{align*}
In our tensor network diagrams, we denote tensors in canonical form using triangles, pointing in the direction of the uncontracted leg. 
 
Similarly, in right canonical form, all tensors (except for the leftmost one) satisfies
\begin{align}
    \sum_{r,u} R_{l,r}(u) \,\,\, & R^*_{l',r}(u) = \delta_{l,l'} % \mathbb{I}\,(l,l')
\end{align}
\begin{align*}
    \begin{tikzpicture}[baseline,scale=0.5]
        % \draw[thick,rounded corners] (-0.5,-0.125) -- (-0.5,2.625) -- (0.75,2.625) -- (0.75,-0.125) -- cycle;
        \draw[thick,rounded corners] (-0.5,0.5) -- (-0.5,2.0);
        \draw[thick,rounded corners] (-1.25,0.5)--(0.5,0.5)--(0.5,2.0)--(-1.25,2.0);
        \node at (-0.2, 1.25) {\scriptsize $u$};
        \node at ( 0.8, 1.25) {\scriptsize $r$};
        \draw[fill=black] (-0.35,2.2)--(-0.75,2.0)--(-0.35,1.8)--cycle;
        \draw[fill=black] (-0.35,0.7)--(-0.75,0.5)--(-0.35,0.3)--cycle;
        \node[rotate=0] at (1.75,1.25) {=};
        \draw[thick,rounded corners] (2.5,0.5)--(3.75,0.5)--(3.75,2.0)--(2.5,2.0);
        \draw[fill=white] (3.25,0.75) rectangle (4.25,1.75);
        \node at (3.75,1.25) {$\mathbb{I}$};
    \end{tikzpicture}
\end{align*}

We are able to define canonical forms because tensor networks have a gauge degree of freedom. This means that the choice of tensors in the network is not unique. One can see this by introducing a set of matrices $X$, $X^{-1}$, which clearly satisfy $XX^{-1}=\mathbb{I}$, along any line connecting two tensors.

Canonicalization can be performed using singular value decomposition (SVD). Suppose that we are interested in written the MPS in left canonical form, and that all tensors left of the $n^\text{th}$ tensor are already in left canonical form. We then take the SVD of the $n^\text{th}$ tensor,
\begin{align}
    F^{(n)}_{c,c'}(i) = \sum_\sigma U_{c,\sigma}(i) \, \Sigma_\sigma V^\dag_{\sigma,c'}
\end{align}
where $\Sigma$ are the singular values from the decomposition of the tensor. Note that by definition, $U$ is left canonical, as desired, and thus will be used as the new $n^\text{th}$ tensor. The remaining matrices are then pushed into the $(n+1)^\text{th}$ tensor,
\begin{align}
    & F^{(n)}_{c,\sigma}(i) \leftarrow U_{c,\sigma}(i) \\ 
    & F^{(n+1)}_{\sigma,c''}(i) \leftarrow \sum_{c'} \Sigma_\sigma V^\dag_{\sigma,c'} F^{(n+1)}_{c',c''}(i) \,\, .
\end{align}
By iteratively performing this operation starting from the left-most tensor all the way to the right end of the MPS, the MPS is put into left canonical form. The procedure for expressing the MPS in right canonical form is analogous. 

MPS compression is performed in the same way, except only the largest $D$ singular values are retained, generating some error. For minimal compression errors, the MPS must be in left (right) canonical form prior to performing the iterative compression procedure starting from the right (left) end.

In Fig.~\ref{diag:IF_alg} we depict the contraction of the columns of the (1+1)D influence functional tensor network. The rows are initialized in left canonical form. After the two right-most columns are contracted, the product is canonicalized and the compressed using the procedure discussed above. Because of the vertical orientation, the left and right canonical forms are depicted by triangles pointing upwards and downwards along the column.

% The tensors have a gauge degree of freedom because over any summed bond, one can insert matrices X, Xdag which can be grouped with adjacent tensors (figure with two tensors, gauge matrices and algebraic expression). The arrows in Fig. X explicitly mean (draw picture, and same equation with labels).

\section{Matrix Product Form of Analytical IF}

As discussed in the text, the expression for the analytical IF in discretized time steps is given by Eq.~\eqref{eq:exactSB} and we wish to write it in the MPS form with physical bonds that index the states of the density matrix at each timestep,
\begin{align}
I(s_{t_1}, & s_{t_2},\ldots, s_{t_{N}}) = \nonumber \\
& \sum_{i_1,...,i_{N-1}}
A(s_{t_1})_{i_1} A(s_{t_2})_{i_1,i_2} \ldots A(s_{t_{N}})_{i_{N-1}}
\label{eq:if_anl}
\end{align}
One possible way to construct the IF is to take the product of factors in Eq.~\eqref{eq:exactSB} in the order $I = I_0  I_1  \ldots I_{N-1}$. 
We start by using the $I_0$ terms which are in the form of a product state (MPS with bond dimension $1$). We then multiply by each of the subsequent $I_m$ and compress into an MPS after each $I_m$ is applied. Multiplying by $I_m$ can be viewed as multiplication by an MPO where the tensors are very sparse.
The two-body terms $I_m$ for $m>1$ are long-range operators, and must be padded with identities to skip over the times in the middle. 
More explicitly, the MPOs are
\begin{align}
    & I_m(s_{t_k}, s_{t_{k+m}}) \, = \sum_v \left[I_m^{(0)} (s_{t_k}) \right]_v  \left[I_m^{(1)} (s_{t_{k+m}}) \right]_v
    \\
    & \hspace{0.5cm} \rightarrow \hspace{0.45cm}
    \begin{tikzpicture}[baseline,scale=0.6]
        \draw[thick] (-4.0, 0) -- (-0.75,0);
        \draw[thick] ( 0.5, 0) -- ( 4.0,0);
        \node at (-0.15,0) {$\hdots$};
        \draw[thick] (-4.0,-1.0) -- (-4.0,1.0);
        \draw[thick] (-2.0,-1.0) -- (-2.0,1.0);
        \draw[thick] ( 2.0,-1.0) -- ( 2.0,1.0);
        \draw[thick] ( 4.0,-1.0) -- ( 4.0,1.0);
        \draw[fill=white] (-4.6,-0.5) rectangle (-3.4,0.5);
        \draw[fill=white] (-2.6,-0.5) rectangle (-1.4,0.5);
        \draw[,fill=white] ( 1.0,-0.5) rectangle ( 3.0,0.5);
        \draw[,fill=white] ( 3.4,-0.5) rectangle ( 4.6,0.5);
        \node at (-4,-1.5) {\scriptsize $s_{t_k}$};
        \node at (-2,-1.5) {\scriptsize $s_{t_{k+1}}$};
        \node at (-0.15,-1.5) {\scriptsize $\hdots$};
        \node at ( 2,-1.5) {\scriptsize $s_{t_{k+m-1}}$};
        \node at ( 4,-1.5) {\scriptsize $s_{t_{k+m}}$};
        \node at (-4, 1.5) {\scriptsize $s'_{t_k}$};
        \node at (-2, 1.5) {\scriptsize $s'_{t_{k+1}}$};
        \node at (-0.15, 1.5) {\scriptsize $\hdots$};
        \node at ( 2, 1.5) {\scriptsize $s'_{t_{k+m-1}}$};
        \node at ( 4, 1.5) {\scriptsize $s'_{t_{k+m}}$};
        \node at (-4.0,0) {\scriptsize $B^{(0)}$};
        \node at (-2.0,0) {\scriptsize $B^{(1)}$};
        \node at ( 2.0,0) {\scriptsize $B^{(m-1)}$};
        \node at ( 4.0,0) {\scriptsize $B^{(m)}$};
    \end{tikzpicture}
    \nonumber
\end{align}
where we first decompose $I_m$ into two tensors (eg. via SVD or QR decomposition) and then define the tensors in the MPO as
\begin{align}
    & B^{(0)}_v (s',s) \hspace{-0.35cm} && = \hspace{0.15cm} \sum_{j} \left( \delta_{s,s'} \delta_{s,v} \right) \left[I^{(0)}_m (s) \right]_{v} \\
    & B^{(m)}_v (s',s) \hspace{-0.35cm} &&= \hspace{0.15cm} \sum_{j} \left( \delta_{s,s'} \delta_{s,v} \right) \left[I^{(1)}_m (s') \right]_{v} \\
    & B^{(i)}_{v,v'} (s',s) \hspace{-0.35cm} &&= \hspace{0.15cm} \delta_{v,v'} \delta_{s,s'} \hspace{0.5cm} \forall \, i \in [1,...,m-1] \,\, .
\end{align}

In diagrammatic form, the MPS for an IF with 6 time steps is
\begin{align*}
\footnotesize
\begin{tikzpicture}[baseline,scale=0.4]
    %\draw[thick] (-5,0) -- ( 5,0);
    \draw[thick] (-5,0) -- (-5,11);
    \draw[thick] (-3,0) -- (-3,11);
    \draw[thick] (-1,0) -- (-1,11);
    \draw[thick] ( 1,0) -- ( 1,11);
    \draw[thick] ( 3,0) -- ( 3,11);
    \draw[thick] ( 5,0) -- ( 5,11);
    \draw[thick,fill=white] (-5,0) circle (0.25);
    \draw[thick,fill=white] (-3,0) circle (0.25);
    \draw[thick,fill=white] (-1,0) circle (0.25);
    \draw[thick,fill=white] ( 1,0) circle (0.25);
    \draw[thick,fill=white] ( 3,0) circle (0.25);
    \draw[thick,fill=white] ( 5,0) circle (0.25);
    \draw[thick,fill=white] (-5.5,1.00) rectangle (-2.5,1.25);
    \draw[thick,fill=white] (-3.5,1.50) rectangle (-0.5,1.75);
    \draw[thick,fill=white] (-1.5,2.00) rectangle ( 1.5,2.25);
    \draw[thick,fill=white] ( 0.5,2.50) rectangle ( 3.5,2.75);
    \draw[thick,fill=white] ( 2.5,3.00) rectangle ( 5.5,3.25);
    \draw[thick,fill=white] ( 0.5,4.00) rectangle ( 5.5,4.25);
    \draw[thick,fill=white] (-1.5,4.50) rectangle ( 3.5,4.75);
    \draw[thick,fill=white] (-3.5,5.00) rectangle ( 1.5,5.25);
    \draw[thick,fill=white] (-5.5,5.50) rectangle (-0.5,5.75);
    \draw[thick,fill=white] (-5.5,6.50) rectangle ( 1.5,6.75);
    \draw[thick,fill=white] (-3.5,7.00) rectangle ( 3.5,7.25);
    \draw[thick,fill=white] (-1.5,7.50) rectangle ( 5.5,7.75);
    \draw[thick,fill=white] (-3.5,8.50) rectangle ( 5.5,8.75);
    \draw[thick,fill=white] (-5.5,9.00) rectangle ( 3.5,9.25);
    \draw[thick,fill=white] (-5.5,10.00) rectangle ( 5.5,10.25);
    \node at (-7,0.00) {$I_0$};
    \node at (-7,2.125) {$I_1$};
    \node at (-7,4.875) {$I_2$};
    \node at (-7,7.125) {$I_3$};
    \node at (-7,8.875) {$I_4$};
    \node at (-7,10.125) {$I_5$};
    \draw[rounded corners] (-6.0,0.5) -- (-6.25,0.5) -- (-6.25,-0.5) -- (-6.0,-0.5);
    \draw (-6.25,0.0) -- (-6.5,0.0);
    \draw[rounded corners] (-6.0,0.75) -- (-6.25,0.75) -- (-6.25,3.5) -- (-6.0,3.5);
    \draw (-6.25,2.125) -- (-6.5,2.125);
    \draw[rounded corners] (-6.0,3.75) -- (-6.25,3.75) -- (-6.25,6.0) -- (-6.0,6.0);
    \draw (-6.25,4.875) -- (-6.5,4.875);
    \draw[rounded corners] (-6.0,6.25) -- (-6.25,6.25) -- (-6.25,8.0) -- (-6.0,8.0);
    \draw (-6.25,7.125) -- (-6.5,7.125);
    \draw[rounded corners] (-6.0,8.25) -- (-6.25,8.25) -- (-6.25,9.5) -- (-6.0,9.5);
    \draw (-6.25,8.875) -- (-6.5,8.875);
    \draw[rounded corners] (-6.0,9.75) -- (-6.25,9.75) -- (-6.25,10.50) -- (-6.0,10.50);
    \draw (-6.25,10.125) -- (-6.5,10.125);
\end{tikzpicture}
\normalsize
\end{align*}
In our contraction scheme, we start from the bottom row and contract upwards. However, because each $I_m$ factor commutes with the rest, other choices of ordering are possible and are the basis of algorithms such as TEMPO~\cite{Strathearn_TEMPO, Jorgensen_TEMPO}.

\section{Influence functional transverse contraction around an arbitrary site}

Sometimes the site whose dynamics we are interested in may be at the middle of the MPS representation of the system (e.g. in the hard core boson model). Thus, we need to generalize the tensor network diagrams presented in the main text to consider IFs for subsystems at arbitrary lattice site $i$. 

\begin{figure}[h]
    % \centering
    % \input{DIAG_si_fig_IFmid}
    \includegraphics[width=\linewidth,trim={0cm 0cm 0cm 0cm}]{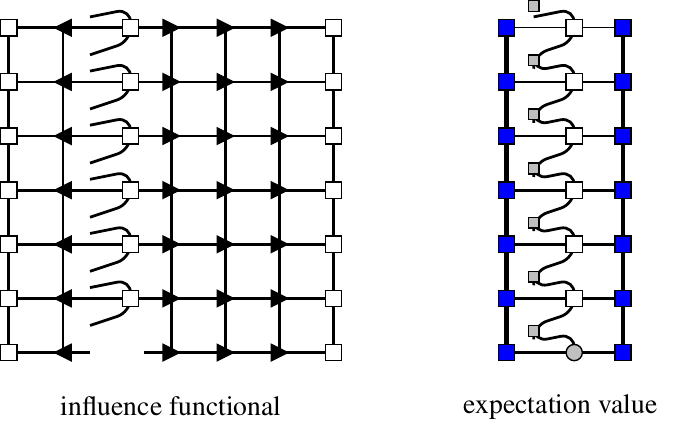}
    \caption{(left) Tensor network showing generalized IF isolating dynamics at the $i=3$ lattice site. Triangles denote gauging of tensors along the row, as defined in Fig.~\ref{diag:IF_alg} in the main text. (right) Tensor network computing expectation value using [blue] left and right environment columns obtained separately using the iterative contraction scheme described above, [white] original tensors at site $i$ dictating interactions of both environment columns with site itself and the environments with each other, and [gray] on-site terms including [circle] the initial state and [square] time evolution operators and the observable of interest (see Fig.~\ref{diag:IF_TN}). It is cheapest to contract this network vertically from the row at one end and continuing to the other end. Note that in using this method one does not explicitly compute the IF itself.}
    \label{diag:mid}
\end{figure}

If the Hamiltonian only consists of nearest-neighbor interactions, the IFs from the two sides of site $i$ are separable and can be computed independently. Otherwise, the tensor network can be initialized as shown in Fig.~\ref{diag:mid}, and one contracts inwards from the outer columns separately. Once only the column corresponding to the isolated site is left, one can now include on-site terms (initial state, on-site time evolution operators, observable) such that the network now corresponds to the expectation value of the observable at the desired time step (a scalar). The cost of contracting this network scales like $\mathcal{O}(D_I^3)$, which is much cheaper than explicitly computing the full IF first and then computing the observable expectation values.

\vfill

\bibliographystyle{apsrev4-1}
\bibliography{main}

\end{document}